\documentclass[doublecolumn]{pasj01}%[doublecolmn]
\Received{2018 December 5}
\Accepted{2019 May 17}
\Published{$\langle$publication date$\rangle$}
\begin{document}

\title{Nobeyama 45 m Mapping Observations toward Orion A. II. Classification of cloud structures and variation of the $^{13}$CO/C$^{18}$O abundance ratio due to far-UV Radiation}
\author{Shun \textsc{Ishii}\altaffilmark{1,2},
Fumitaka \textsc{Nakamura}\altaffilmark{1,3,4},
Yoshito \textsc{Shimajiri}\altaffilmark{1,5,6},
Ryohei \textsc{Kawabe}\altaffilmark{1,3,4},
Takashi \textsc{Tsukagoshi}\altaffilmark{1},
Kazuhito \textsc{Dobashi}\altaffilmark{7}, and 
Tomomi \textsc{Shimoikura}\altaffilmark{7,8}}

\altaffiltext{1}{National Astronomical Observatory of Japan, 2-21-1 Osawa, Mitaka, Tokyo 181-8588, Japan}
\altaffiltext{2}{Joint ALMA Observatory, Alonso de C\'ordova 3107 Vitacura, Santiago, Chile}
\altaffiltext{3}{The Graduate University for Advanced Studies (SOKENDAI), 2-21-1 Osawa, Mitaka, Tokyo 181-0015, Japan}
\altaffiltext{4}{The University of Tokyo, 7-3-1 Hongo Bunkyo, 113-0033 Tokyo, Japan}
%\altaffiltext{4}{Nobeyama Radio Observatory, National Astronomical Observatory of Japan 462-2 Nobeyama, Minamimaki, Minamisaku, Nagano 384-1305}
\altaffiltext{5}{Laboratoire AIM, CEA/DSM-CNRS-Universit\'e Paris Diderot, IRFU/Service d' Astrophysique, CEA Saclay, 91191 Gif-sur-Yvette, France}
\altaffiltext{6}{Department of Physics and Astronomy, Graduate School of Science and Engineering, Kagoshima University, 1-21-35 Korimoto, Kagoshima, Kagoshima 890-0065, Japan}
\altaffiltext{7}{Department of Astronomy and Earth Sciences, Tokyo Gakugei University, 4-1-1 Nukuikitamachi, Koganei, Tokyo 184-8501, Japan}
\altaffiltext{8}{Otsuma Women's University Faculty of Social Information Studies Chiyoda-ku, Tokyo 102-8357, Japan}
\email{shun.ishii@nao.ac.jp}


\maketitle

\begin{abstract}
We present results of the classification of cloud structures toward the Orion A Giant Molecular Cloud based on wide-field $^{12}$CO ($J$ = 1--0), $^{13}$CO ($J$ = 1--0), and C$^{18}$O ($J$ = 1--0) observations using the Nobeyama 45 m radio telescope. 
We identified 78 clouds toward Orion A by applying Spectral Clustering for Interstellar Molecular Emission Segmentation (SCIMES) to the data cube of the column density of $^{13}$CO.
Well-known subregions such as OMC-1, OMC-2/3, OMC-4, OMC-5, NGC 1977, L1641-N, and the dark lane south filament (DLSF) are naturally identified as distinct structures in Orion A.
These clouds can also be classified into three groups: the integral-shaped filament, the southern regions of Orion A, and the other filamentary structures in the outer parts of Orion A and the DLSF.
These groups show differences in scaling relations between the physical properties of the clouds.
We derived the abundance ratio between $^{13}$CO and C$^{18}$O, $X_{^{13}\mathrm{CO}}/X_{\mathrm{C}^{18}\mathrm{O}}$, which ranges from 5.6 to 17.4 on median over the individual clouds. The significant variation of $X_{^{13}\mathrm{CO}}/X_{\mathrm{C}^{18}\mathrm{O}}$ is also seen within a cloud in both of the spatial and velocity directions and the ratio tends to be high at the edge of the cloud. 
The values of $X_{^{13}\mathrm{CO}}/X_{\mathrm{C}^{18}\mathrm{O}}$ decrease from 17 to 10 with the median of the column densities of the clouds at the column density of $N_{\mathrm{C^{18}O}} \gtrsim 1 \times 10^{15}$ cm$^{-2}$ or visual extinction of $A_V \gtrsim 3$ mag under the strong far-ultraviolet (FUV) environment of $G_0 > 10^{3}$, whereas it is almost independent of the column density in the weak FUV radiation field.  
These results are explained if the selective photodissociation of C$^{18}$O is enhanced under a strong FUV environment and it is suppressed in the dense part of the clouds.
\end{abstract}

\section{Introduction}
\label{sec:intro}
The process of star formation is deeply connected with the structures of giant molecular clouds (GMCs).
GMCs are composed of hierarchical structures: small-scale dense cores are contained inside filaments that are enveloped in lower density gas \citep{Rosolowsky08}.
Over the wide scale of these structures, self-gravity plays an important role in leading star formation \citep{Goodman09}.
However, its role in individual star-forming regions remains to be elucidated partly because of a lack of sufficient observational data and  the difficulty in the unbiased identification of cloud structures.
In addition to self-gravity, stellar feedback such as protostellar H{\scriptsize II} regions, outflows, and stellar radiation also significantly influence cloud structure and star formation therein.

The far-ultraviolet (FUV: 6 eV $<h \nu<$ 13.6 eV) radiation from massive stars affects the thermal balance, structure, chemistry, and evolution of neutral interstellar medium \citep{Hollenbach99}. 
Furthermore, the cloud surfaces tend to be photodissociated in the presence of FUV.
In such photon-dominated regions (PDRs) rarer isotopologues of CO can be dissociated deeper inside the clouds due to the difference in self-shielding effects by up to 1--2 orders of magnitude in the photodissociation rate \citep{vanDishoeck88, Visser09}. 
Since the self-shielding is more effective for abundant species, C$^{18}$O is dissociated more than $^{13}$CO and the abundance ratio of $^{13}$CO and C$^{18}$O is likely to become larger under the influence of FUV radiation.  
This isotope-selective effect is called selective photodissociation. Thus, the spatial variation of the $^{13}$CO/C$^{18}$O abundance ratio, $X_{^{13}\mathrm{CO}}/X_{\mathrm{C}^{18}\mathrm{O}}$,  is expected to reveal the area where the FUV radiation affects the cloud structure and properties.
The effect of selective photodissociation is observed in several star-forming regions \citep{Bally82, Lada94, Kong15, Lin16, Paron18}.
\citet{Shimajiri14} revealed the distribution of $X_{^{13}\mathrm{CO}}/X_{\mathrm{C}^{18}\mathrm{O}}$ toward the northern part of the Orion A GMC. They found that $X_{^{13}\mathrm{CO}}/X_{\mathrm{C}^{18}\mathrm{O}}$ values are higher than the solar system value throughout the region.
In particular, the abundance ratio in the nearly edge-on PDRs are $\sim$ 3 times as large. The difference of the ratio is most likely due to the selective FUV photodissociation of C$^{18}$O. Enhancement of the abundance of $^{13}$CO due to the chemical fractionation, $\mathrm{^{13}C^{+} + ^{12}CO \rightarrow ^{13}CO + ^{12}C^{+} + 35 K}$ \citep{Langer80}, was also discussed by \citet{Shimajiri14}. However, in the Orion-A GMC, the temperature is high ($T_{\mathrm{ex}} = 28.4 \pm 9.7$ K) in regions with low column density and thus this isotopic exchange reaction would not be dominant.

The structures within Orion A have been recognized along with the progress of millimeter, submillimeter, and infrared observations. Orion molecular cloud 1 (OMC-1) was identified as a dense gas directly associated with Orion KL \citep{Wilson70, Zuckerman73, Liszt74}, then OMC-2 \citep{Gatley74} and OMC-3 \citep{Kutner76} were detected as subsequent condensations in CO emission located about \timeform{15'} and \timeform{25'} to the north of OMC-1. The $^{13}$CO ($J$ =1--0 observations by \citet{Bally87} revealed that these clouds consist of the integral-shaped filament. After that, the SCUBA maps at 450 and 850 $\mu$m presented concentrations of submillimeter continuum emission in the southern part of the integral-shaped filament, which are now referred to as OMC-4 \citep{Johnstone99} and OMC-5 \citep{Johnstone06}.
In parallel with these observations, the distribution of molecular gas in Orion A GMC has been quite extensively studied in multiple rotational transition lines of CO and its isotopologues: $^{12}$CO ($J$ =1--0) \citep{Maddalena86, Wilson05, Shimajiri11, Nakamura12}, $^{12}$CO ($J$ =1--0) and $^{13}$CO ($J$ =1--0) \citep{Ripple13}, $^{12}$CO ($J$ =2--1) and $^{13}$CO ($J$ =2--1) \citep{Berne14},  $^{12}$CO ($J$ =2--1), $^{13}$CO ($J$ =2--1) and C$^{18}$O($J$ =2--1) \citep{Nishimura15}, $^{12}$CO ($J$ = 2--1) \citep{Sakamoto94}, $^{12}$CO ($J$ = 3--2) \citep{Takahashi08}, $^{12}$CO ($J$ = 4--3) \citep{Ishii16}. $^{13}$CO ($J$ = 1--0) \citep{Nagahama98}, $^{13}$CO ($J$ = 1--0) and C$^{18}$O ($J$ = 1--0) \citep{Shimajiri14}, and $^{13}$CO ($J$ = 3--2) and C$^{18}$O ($J$ = 3--2) \citep{Buckle12}.

This paper investigates how the FUV radiation affects individual clouds in a GMC by combining the results of the identification of cloud structure and estimation of the abundances of CO isotopologues, $^{13}$CO and C$^{18}$O in three-dimensional space. 
The target is the nearest high-mass star-forming giant molecular cloud, Orion A ($d = 414 \pm 7$ pc by \cite{Menten07}), which are newly observed over a 7 pc $\times$ 15 pc region.
This wide-field observation allows us to examine the relation between cloud properties and the FUV radiation in an extensive range of field strength, i.e., from a strong FUV ($G_0 \sim 10^{4-5}$ in units of the local interstellar radiation field by \cite{Habing68}) environment in the northern part to a weaker ($G_0 \sim 10^{1}$) environment in the southern part of the GMC. 
Following a description of the observations and data sets analyzed here in section 2, we derive distributions of excitation temperature, optical depth, and column density and provide an overview of structures within Orion A in section 3. The three-dimensional distribution of the abundance ratio of $^{13}$CO and C$^{18}$O is also estimated. In section 4, the details of the identification of molecular clouds and clustering analysis by SCIMES and the results are reported. In section 5, we discuss the interpretation of the results of the clustering analysis and scaling relations between physical properties such as size, line width, mass, and virial parameter. Finally, we examine the effect of the selective photodissociation of C$^{18}$O for defined clouds.

The present observations of Orion A were made as a part of the Nobeyama Star Formation Legacy Project at the Nobeyama Radio Observatory (NRO) to observe nearby star-forming regions, Orion A, Aquila Rift, and M17. 
The overview of the project will be presented in a separate paper \citep{Nakamura19a} and the detailed observational results for the individual regions are given in other articles 
(Orion A: \cite{Tanabe19}, \cite{Nakamura19b}, Aquila Rift: \cite{Shimoikura19b}, \cite{Kusune19}, M17: \cite{Shimoikura19a}, other regions: \cite{Dobashi19a,Dobashi19b}).

\section{Observations and Data}
\label{sec:obs}

\subsection{Observations with the Nobeyama 45 m radio telescope}
We carried out observations in the emission lines of $^{12}$CO ($J$ = 1--0), $^{13}$ CO ($J$ = 1--0),  and C$^{18}$O ($J$ = 1--0) toward the Orion A GMC using the 45 m radio telescope of the NRO. 
The $\timeform{2.0D} \times \timeform{1.5D}$ region in Orion A was simultaneously mapped in $^{12}$CO ($J$ = 1--0) and $^{13}$ CO ($J$ = 1--0) in 2015 May and the period from 2015 December to 2016 May. 
The observations in C$^{18}$O ($J$ = 1--0) were done with the same observing method in the period from 2016 December to 2017 March.  
The telescope has a beam size of \timeform{14.1"} (half-power beamwidth) at 115 GHz.
We used a new four-beam receiver, FOREST, which is a dual-polarization sideband-separating SIS receiver 
\citep{Minamidani16}. 
FOREST has 16 intermediate frequency (IF) outputs in total; eight IFs in the upper-sideband and the other eight IFs in the lower-sideband were used.
The beam separation of FOREST is $\timeform{51.7"}$ on the sky.  
As the receiver backend, we used a digital spectrometer based on an FX-type correlator, SAM45, that has 16 sets of 4096-channel arrays.
In these observations, the total bandwidth and frequency resolution of all spectrometer arrays were set to 62.5 MHz and 15.26 kHz, respectively, which correspond to 163 km s$^{-1}$ and 0.04 km s$^{-1}$ at 115 GHz.
The observed region was covered by a combination of small observation boxes with a size of $\timeform{10'}  \times \timeform{10'}$.
The on-the-fly (OTF) technique \citep{Sawada08} was adopted to map each observation box.
Scans of the OTF observations are separated by \timeform{5''.17}, so that individual scans by the four beams of FOREST are overlapped. 

As an ``off'' position, we used an emission-free area at ($\alpha_{\mathrm{J2000.0}}, \delta_{\mathrm{J2000.0}}) = (\timeform{5h29m00.0s}, \timeform{-5D25'30.0''}$) that is $\sim \timeform{2D}$ away from the mapping area. 
The typical system noise temperature was in the range from 350 K to 400 K for $^{12}$CO ($J$ = 1--0) and from 150 K to 200 K for $^{13}$CO ($J$ = 1--0) and C$^{18}$O ($J$ = 1--0) in the single sideband mode at the observed elevation El=\timeform{30D}--\timeform{50D}.
The temperature scale was determined by the chopper-wheel method.
The telescope pointing was checked every 1 hour by observing the SiO maser line from Orion KL ($\alpha_{\mathrm{J2000.0}}, \delta_{\mathrm{J2000.0}} = \timeform{5h35m14.5s}, \timeform{-5D22'30.4''}$). The pointing accuracy was better than $\sim$ 5$''$ throughout the observations.
The parameters of observations are summarized in table \ref{tab:obs}.
In order to minimize the scanning effects, the data with orthogonal scanning directions along the right ascension and declination axes were combined into a single map. 
We adopted a spheroidal function as a gridding convolution function to calculate the intensity at each grid point of the final cube data with a spatial grid size of \timeform{7.5"}. The final effective angular resolutions are \timeform{21.6"} for $^{12}$CO ($J$ = 1--0) and \timeform{22.0"} for $^{13}$CO ($J$ = 1--0) and C$^{18}$O ($J$ = 1--0), respectively. 
The spheroidal function used in the NRO reduction software is explained by \citet{Sawada08} and the function itself is described by \citet{Schwab84} in detail.   
We applied the parameters $m = 6$ and $\alpha = 1$, which define the shape of the function.
The intensity scale of the data was calibrated by comparing with the previous survey data taken by the BEARS receiver \citep{Shimajiri11, Nakamura12, Shimajiri14}, which was already converted to $T_{\rm mb}$ the brightness temperature scale corrected for the main beam efficiency.
We determine a scale factor for the data of each observation box and spectrometer array.
We produced the integrated intensity maps of the FOREST and BEARS data after matching the grid of the FOREST data to that of the BEARS data.
We then defined the scale factor as the ratio of the mean integrated intensities of the FOREST and BEARS maps.
The scale factors of the arrays range between 2.0 -- 2.6 for $^{12}$CO ($J$ = 1--0) and between 2.7 -- 3.2 for both $^{13}$CO ($J$ = 1--0) and C$^{18}$O ($J$ = 1--0) .
The velocity resolutions of the final datasets after merging the FOREST and BEARS data are 0.198 km s$^{-1}$ for the $^{12}$CO ($J$= 1--0) data, and 0.220 km s$^{-1}$ for the $^{13}$CO ($J$= 1--0) and C$^{18}$O ($J$= 1--0) data.
The noise levels were $\Delta T_{\rm mb} = 0.48  \pm 0.08$ K  for $^{12}$CO data, $\Delta T_{\rm mb} = 0.18 \pm 0.04$ K for $^{13}$CO data, and $\Delta T_{\rm mb} = 0.24 \pm 0.03$ K for C$^{18}$CO data.  See Nakamura et al. (2019a) for more details.

\begin{table*}
  \tbl{Observed lines}{%
  \begin{tabular}{llllll}
  \hline
  Molecule &Transition & Rest frequency &Beam size & Velocity resolution & Noise level \\
& &(GHz)  & (arcsec) & (km s$^{-1}$) & (K) \\ 
\hline
$^{12}$CO	& $J$=1--0 &	115.271204  	& \timeform{21.6} 	& 0.198 & 0.48 $\pm$ 0.08 \\
$^{13}$CO	& $J$=1--0 & 	110.201354 	& \timeform{22.0}	& 0.220 & 0.18 $\pm$ 0.04  \\
C$^{18}$O	& $J$=1--0 & 	109.782170	& \timeform{22.0}	& 0.220 & 0.24 $\pm$ 0.03  \\
\hline
  \end{tabular}}\label{tab:obs}
  \begin{tabnote}
The last column is the average root-mean-square (rms) noise levels of the whole area in the $T_{\rm mb}$ scale. 
The rms noise level is measured in each observation box and is indicated with the standard deviation.
  \end{tabnote}
\end{table*}

\subsection{Dust continuum data}
\label{sec:dust}
In order to examine relations among cloud properties and to probe into the environmental effects in Orion A, we compare the dataset of CO lines with the dust continuum observations from the {\it Herschel} Gould Belt Survey \citep{Andre10} and the {\it Planck} satellite \citep{Planck11}.
Based on the map of the optical depth at 850 $\mu$m and the dust temperature derived by fitting of the spectral energy distribution \citep{Lombardi14}, we converted $\tau_\mathrm{850 \mu m}$ to visual extinction ($A_\mathrm{V}$) using $A_\mathrm{V} = A_\mathrm{K}/0.112$ \citep{Rieke85} and $A_K = \gamma_\mathrm{Orion\ A} \tau_\mathrm{850 \mu m} + \delta_\mathrm{Orion\ A}$, where $\gamma_\mathrm{Orion\ A}$ and $\delta_\mathrm{Orion\ A}$ are estimated to be $2640$ mag and $0.012$ mag, respectively.
The photometric data at $70$ $\mu$m taken by PACS on {\it Herschel} is also taken from the science archive (obsid: 1342218967).
The pixel scale of the original data is \timeform{3.2''}. The bandwidth of the $70$ $\mu$m band is $\Delta \lambda = 10.6$ $\mu$m.
The map of the FUV radiation field $G_0$ can be estimated by following equations (9) and (10) in \citet{Shimajiri17}.
Note that we use only the $70$ $\mu$m map to estimate $G_0$ because the $100$ $\mu$m map is currently not available.  

%SI added 20180528
\section{Results}
\label{sec:results}
\subsection{Excitation temperature, optical depth, and column density}

\begin{figure*}[tbp]
\begin{center}
\FigureFile(150mm,150mm){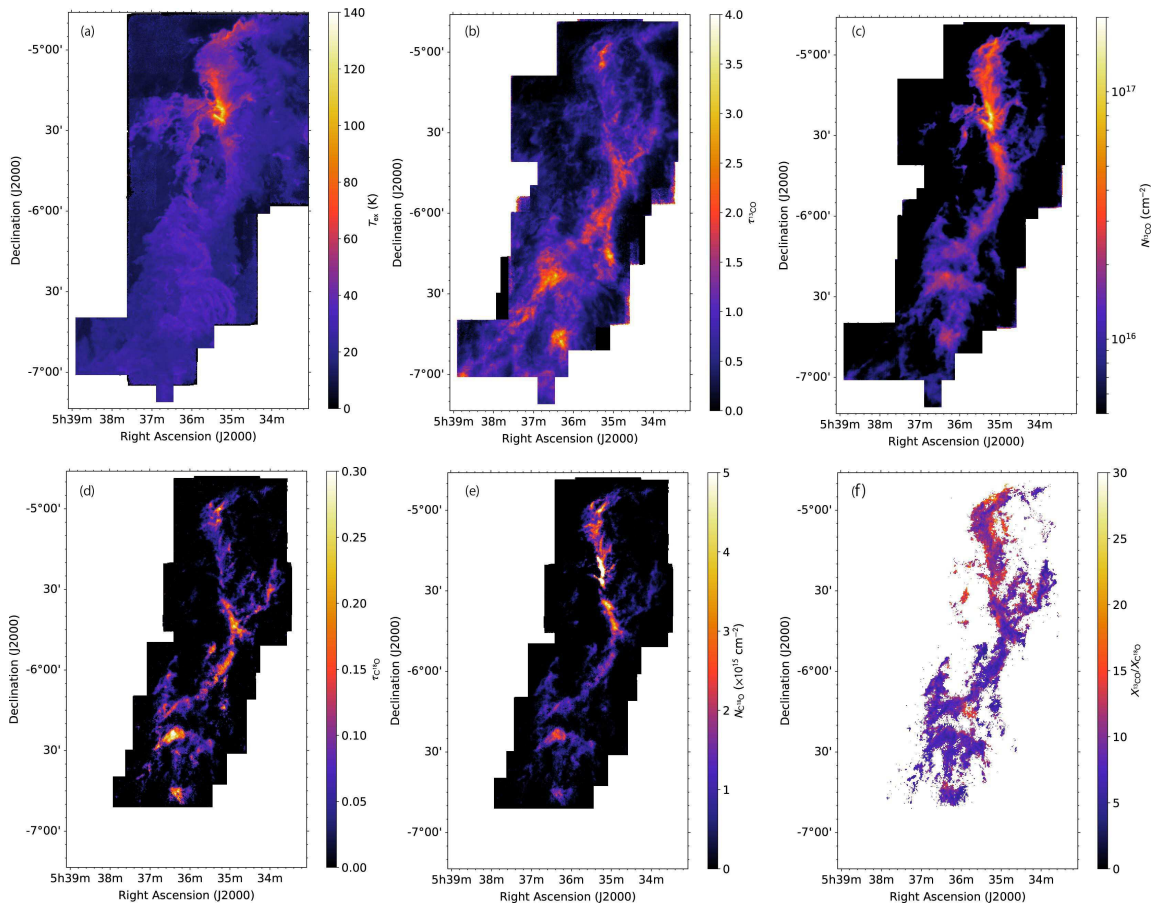}
\vspace{5mm}
\caption{Maps of (a) the excitation temperature derived from the peak intensity of $^{12}$CO ($J$=1--0), (b) the optical depth and (c) the column density of $^{13}$CO ($J$=1--0), (d) the optical depth and (e) the column density of C$^{18}$O($J$=1--0), and (f) the ratio of the fractional abundances of $^{13}$CO and C$^{18}$O,  $X_{^{13}\mathrm{CO}}/X_{\mathrm{C}^{18}\mathrm{O}}$.}
\label{fig:COmaps}
\end{center}
\end{figure*}

We derived the excitation temperature, the optical depth, and the column density across the Orion A GMC using the datasets of $^{12}$CO ($J$=1--0), $^{13}$CO ($J$=1--0), and C$^{18}$O($J$=1--0) emission lines with the assumption that the rotational levels of gas are in the local thermodynamic equilibrium (LTE).
Figure \ref{fig:COmaps}a shows the spatial  distribution of the excitation temperature $T_{\mathrm{ex}}$ derived on the assumption that the $^{12}$CO ($J$=1--0) is optically thick.
$T_{\mathrm{ex}}$ is given by
\begin{equation}
T_{\mathrm{ex}} = \frac{5.53}{\ln{\left\{1 + 5.53 / (T_{\mathrm{peak}} + 0.819) \right\}}} \ \ \mathrm{K}
\end{equation}
where $T_{\mathrm{peak}}$ is the peak intensity of $^{12}$CO ($J$=1--0) in Kelvin.
The average of $T_{\mathrm{ex}}$ over the cloud is 23.7 K with the highest value of 130.0 K at Orion KL. 
The gas in the northern part of the cloud including OMC-1/2/3, Orion bar, and the dark lane south filament (DLSF, known to be a PDR from observations by \cite{Rodriguez-Franco98} and \cite{Shimajiri13}) regions has high excitation temperatures up to $130$ K. 
The recent result from the CARMA--NRO Orion Survey \citep{Kong18} indicates that gas localizes with higher excitation temperatures up to $220$ K around the OMC-1/bar region on the scale of \timeform{8''}.   
The southern part including the L1641-N cluster \citep{Allen08} shows a uniform distribution of the excitation temperature, which does not exceed $50$ K. Most of the gas at relatively high temperatures of 40 -- 50 K in this area is associated with shell structures in L1641-N identified by \citet{Nakamura12}.

Assuming that the $^{13}$CO ($J$=1--0) and C$^{18}$O ($J$=1--0) emission lines are optically thin, the optical depths and column densities of these lines are determined for each velocity channel in the data cubes based on equations in \citet{Shimajiri14}.
The data cubes of both lines cover the velocity range $ 2 < V_{\mathrm{LSR}} < 20$ km s$^{-1}$ with a channel width of 0.22 km s$^{-1}$.  
We re-gridded the data cubes to match a grid spacing of the map of $T_{\mathrm{ex}}$ for this analysis.
The analysis is applied only to pixels having a signal-to-noise ratio larger than 5.0 for each data.
The three-dimensional optical depth of $^{13}$CO ($J$=1--0), $\Delta \tau_{^{13}\mathrm{CO}}(\alpha, \delta, V)$, and the three-dimensional column density of $^{13}$CO, $\Delta N_{^{13}\mathrm{CO}}(\alpha, \delta, V)$, are given by
\begin{equation}
\Delta \tau_{^{13}\mathrm{CO}}(\alpha, \delta, V) = -\ln{ \left\{ 1- \frac{T_{^{13}\mathrm{CO}}/\phi_{^{13}\mathrm{CO}} }{ 5.29[J(T_{\mathrm{ex}}) -0.164]} \right\} },
\end{equation}
and 
\begin{equation}
\Delta N_{^{13}\mathrm{CO}}(\alpha, \delta, V) = 2.42 \times 10^{14} \left\{ \frac{\Delta \tau_{^{13}\mathrm{CO}} \Delta V T_{\mathrm{ex}} }{ 1-\exp{[-5.29/T_{\mathrm{ex}}]}}  \right\} 
\ \ \mathrm{cm^{-2}}
\end{equation}
where $\alpha$, $\delta$, and $V$ are pixels and the channel correspond to right acension, declination, and the velocity, respectively. 
$T_{^{13}\mathrm{CO}}$ is the main beam temperature of the emission at the channel, $\Delta V$ is the velocity width of a channel, and $J(T_{\mathrm{ex}})=1/[\exp{(5.29/T_{\mathrm{ex}})-1}]$.
Similarly, the three-dimensional optical depth of C$^{18}$O ($J$=1--0), $\Delta \tau_{\mathrm{C}^{18}\mathrm{O}}(\alpha, \delta, V)$, and the three-dimensional column density of C$^{18}$O, $\Delta N_{\mathrm{C}^{18}\mathrm{O}}(\alpha, \delta, V)$, are given by
\begin{equation}
\Delta \tau_{\mathrm{C}^{18}\mathrm{O}}(\alpha, \delta, V) = -\ln{ \left\{ 1- \frac{T_{\mathrm{C}^{18}\mathrm{O}}/\phi_{\mathrm{C}^{18}\mathrm{O}} }{ 5.27[J(T_{\mathrm{ex}}) -0.167]} \right\} },
\end{equation}
and 
\begin{equation}
\Delta N_{\mathrm{C}^{18}\mathrm{O}} (\alpha, \delta, V)= 2.42 \times 10^{14} \left\{ \frac{\Delta \tau_{\mathrm{C}^{18}\mathrm{O}} \Delta V T_{\mathrm{ex}} }{ 1-\exp{[-5.27/T_{\mathrm{ex}}]}}  \right\} 
\ \ \mathrm{cm^{-2}}
\end{equation}
where $T_{\mathrm{C}^{18}\mathrm{O}}$ is the main beam temperature of the emission at the channel and $J(T)=1/[\exp{(5.27/T)-1}]$.
The beam filling factors of $\phi_{^{13}\mathrm{CO}}$ and $\phi_{\mathrm{C}^{18}\mathrm{O}}$ for $^{13}$CO ($J$=1--0) and C$^{18}$O ($J$=1--0) are assumed to be 1.0.
The total optical depths and column densities $\tau_{^{13}\mathrm{CO}}(\alpha, \delta)$, $N_{^{13}\mathrm{CO}}(\alpha, \delta)$, $\tau_{\mathrm{C^{18}O}}(\alpha, \delta)$, and $N_{\mathrm{C^{18}O}}(\alpha, \delta)$ can be obtained by integrating these parameters along the velocity axis.

Figures \ref{fig:COmaps}b -- \ref{fig:COmaps}e show spatial distributions of $\tau_{^{13}\mathrm{CO}}(\alpha, \delta)$, $N_{^{13}\mathrm{CO}}(\alpha, \delta)$, $\tau_{\mathrm{C^{18}O}}(\alpha, \delta)$, and $N_{\mathrm{C^{18}O}}(\alpha, \delta)$.
The mean and maximum values of optical depths and the column densities are summarized in table \ref{tab:tau_N}.
The distributions of the optical depth and column density of $^{13}$CO and C$^{18}$O are similar to each other. 
The values of the optical depths are high in the northern edge of OMC-3, OMC-4/5, and L1641-N regions in both emission lines.
The optical depth of $^{13}$CO integrated along the velocity axis exceeds 1.0 in these regions, however the mean value of the optical depth per velocity channel is $0.21 \pm 0.18$, which means that emission of $^{13}$CO ($J$=1--0) can be assumed to be optically thin in the position-position-velocity space. 
The distributions of the column density show that $^{13}$CO and C$^{18}$O are highly concentrated in the integral-shaped filament and L1641-N, with $N_{^{13}\mathrm{CO}}(\alpha, \delta) \gtrsim 5 \times 10^{16}$ cm$^{-2}$ and $N_{\mathrm{C}^{18}\mathrm{O}}(\alpha, \delta) \gtrsim 3 \times 10^{15}$ cm$^{-2}$.  
These distributions also clearly exhibit that the dense components can be identified as sub-filamentary structures in the integral-shaped filament.
As revealed by previous $^{13}$CO maps over Orion A (e.g., \cite{Bally87, Castets90, Nagahama98, Berne14}),  a number of clumpy and filamentary structures are seen in the distribution of $N_{^{13}\mathrm{CO}}(\alpha, \delta)$.
In the northern part of Orion A, these condensations would be separated into structures associated with the integral-shaped filament and others that are located in the eastern or western sides of the filament such as the DLSF and the bending structure \citep{Shimajiri11}.
In contrast, the structures of condensation in the southern part of the map are not elongated compared with filamentary structures in the northern part of Orion A. 
The integral-shaped filament extending from OMC-4/5 is connected to a cloud structure at $\delta_\mathrm{J2000.0} = \timeform{-6D12'}$.
The L1641-N cluster is located at the position of ($\alpha_\mathrm{J2000.0}$, $\delta_\mathrm{J2000.0}$) $=$ (\timeform{5h36m18s},  \timeform{-6D21'48''}) \citep{Nakamura12} and is contained in an extended cloud between $\delta_\mathrm{J2000.0}= \timeform{-6D20'}$ and $\timeform{-6D30'}$.   
Further south, there is a cloud associated with V380 Ori at ($\alpha_\mathrm{J2000.0}$, $\delta_\mathrm{J2000.0}$) $=$ (\timeform{5h36m25.43s},  \timeform{-6D42'57.7''}). 
This cloud has uniform values of $N_{^{13}\mathrm{CO}}(\alpha, \delta) \lesssim 2 \times 10^{16}$ cm$^{-2}$.  

\begin{table*}[tbp]
  \tbl{Mean and maximum values of optical depths and column densities}{%
  \begin{tabular}{llllll}
  \hline
Molecule	&Transition 	& $\tau_{\mathrm{mean}}$ 	& $\tau_{\mathrm{max}}$ 	& $N_{\mathrm{mean}}$ & $N_{\mathrm{max}}$  \\
		&			&						& 						& (cm$^{-2}$) 			&  (cm$^{-2}$) 		\\ 
\hline
$^{13}$CO& $J$=1--0 	& 	0.82					& 3.60					& $6.6 \times10^{15}$	& $1.7 \times10^{17}$  \\
C$^{18}$O& $J$=1--0 	& 	$5.7\times 10^{-2}$		& 0.42					& $7.7 \times 10^{14}$	& $1.7 \times10^{16}$  \\
\hline
  \end{tabular}}\label{tab:tau_N}
  \begin{tabnote}
The mean value is calculated with pixels having signal-to-noise ratio larger than 5.0 for each data.
  \end{tabnote}
\end{table*}

\subsection{Abundance ratio of $^{13}$CO and C$^{18}$O}

\label{sec:abundance}
\begin{figure*}[tbp]
\begin{center}
\FigureFile(120mm,200mm){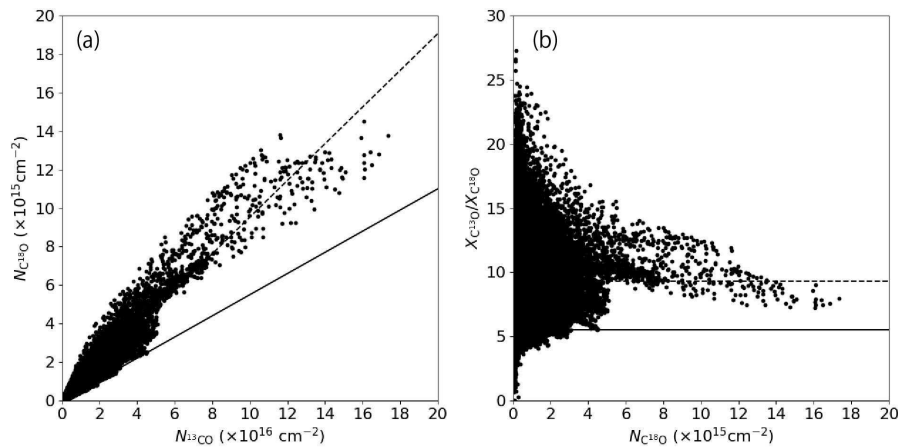}
\vspace{5mm}
\caption{(a) Correlation between $N_{^{13}\mathrm{CO}}$ and $N_{\mathrm{C^{18}O}}$ over the mapped area, and (b) Relation between $N_{\mathrm{C}^{18}\mathrm{O}}$ and $X_{^{13}\mathrm{CO}}/X_{\mathrm{C}^{18}\mathrm{O}}$ over the mapped area. The solid and dashed lines indicate $X_{^{13}\mathrm{CO}}/X_{\mathrm{C}^{18}\mathrm{O}} = 5.5$ and $9.31$.}
\label{fig:x13x18}
\end{center}
\end{figure*}

Because the fractional abundances are proportional to their column densities, the derived column densities of $^{13}$CO and C$^{18}$O allow us to estimate their abundance ratio as follows,
\begin{equation}
\frac{X_{^{13}\mathrm{CO}}(\alpha, \delta)}{X_{\mathrm{C}^{18}\mathrm{O}}(\alpha, \delta)} = \frac{N_{^{13}\mathrm{CO}(\alpha, \delta)}/N_{\mathrm{H}_2}(\alpha, \delta)}{N_{\mathrm{C}^{18}\mathrm{O}(\alpha, \delta)}/N_{\mathrm{H}_2}(\alpha, \delta)} = \frac{N_{^{13}\mathrm{CO}}(\alpha, \delta)}{N_{\mathrm{C}^{18}\mathrm{O}}(\alpha, \delta)}.
\label{eq:x1318}
\end{equation}
Figure \ref{fig:COmaps}f shows the map of $X_{^{13}\mathrm{CO}(\alpha, \delta)}/X_{\mathrm{C}^{18}\mathrm{O}(\alpha, \delta)}$. The distribution of the abundance ratio in the northern part of Orion A is consistent with the results reported by \citet{Shimajiri14}. 
We found that the ratio of the fractional abundances is well consistent with the mean value of $9.31 \pm 2.65$ in the observed area. This is lower than the mean abundance ratio for the northern part of Orion A due to the small value of the ratio in the southern part of the cloud. The abundance ratio significantly varies from $\sim 5$ to $27.3$  in the map (see correlation plots shown in figures \ref{fig:x13x18}a and \ref{fig:x13x18}b.  This implies that the abundance ratio is affected by environmental differences such as the field of the UV radiation and cloud structures for each location in the Orion A GMC.

The abundance ratio can be extended to three-dimensions on the analogy of equation (\ref{eq:x1318}) using the data cubes of column densities of $^{13}$CO and C$^{18}$O as,
\begin{equation}
\frac{\Delta X_{^{13}\mathrm{CO}}(\alpha, \delta, V)}{\Delta X_{\mathrm{C}^{18}\mathrm{O}}(\alpha, \delta, V)} = \frac{\Delta N_{^{13}\mathrm{CO}}(\alpha, \delta, V)}{\Delta N_{\mathrm{C}^{18}\mathrm{O}}(\alpha, \delta, V)}.
\end{equation}
The three-dimensional abundance ratio allows us to investigate the distribution of the abundance ratio in the position-position-velocity space. 
The mean value of $\Delta X_{^{13}\mathrm{CO}}(\alpha, \delta, V)/\Delta X_{\mathrm{C}^{18}\mathrm{O}}(\alpha, \delta, V)$ are found to be $9.53 \pm 3.00$ with the maximum of 35.7. 
The large mean and maximum values compared to those in the two-dimensional ratio means that the abundance ratio also varies in the velocity space. 
The highest value of $\Delta X_{^{13}\mathrm{CO}}/\Delta X_{\mathrm{C}^{18}\mathrm{O}}$ is detected in the northern edge of OMC-2/3 at ($\alpha_{\mathrm{J2000.0}}, \delta_{\mathrm{J2000.0}}, V_{\mathrm{LSR}}$) = (\timeform{5h35m25s}, \timeform{-4D57'15.9''}, 12.4 km s$^{-1}$). 
The value of $X_{^{13}\mathrm{CO}}/X_{\mathrm{C}^{18}\mathrm{O}}$ is also high ($= 21.6$) in this region.
In general, the redshifted components of gas tend to show high $\Delta X_{^{13}\mathrm{CO}}/\Delta X_{\mathrm{C}^{18}\mathrm{O}}$ in our data set.
The velocity components between 8.0 and 10.0 km s$^{-1}$ in the southern part of Orion A ($\delta < \timeform{-5D40'}$) shows $\Delta X_{^{13}\mathrm{CO}}/\Delta X_{\mathrm{C}^{18}\mathrm{O}} > 10$. 

\section{Identification of molecular clouds}
\label{sec:global}

\subsection{Clustering analysis by SCIMES}

We identify internal cloud structures that compose Orion A.
\citet{Colombo15} designed SCIMES (Spectral Clustering for Interstellar Molecular Emission Segmentation) for the analysis of cloud structures based on graph theory and cluster analysis. 
SCIMES offers a novel and robust approach to finding discrete regions with similar emission properties using the hierarchical structure of molecular gas extracted by dendrogram \citep{Rosolowsky08}. 
$^{13}$CO is an appropriate tracer to analyze internal structures through entire molecular clouds. 
$^{12}$CO is optically thick and is a good probe to estimate the kinematic temperature near the surface of the cloud. 
As we derived in section \ref{sec:results}, the isotopolog $^{13}$CO is optically thin in most directions and its emission is a probe of the column density. 
It is demonstrated that the integrated intensity of the $^{13}$CO emission shows a good correlation with $A_V$ \citep{Nakamura17}.
The emission of C$^{18}$O is also optically thin and therefore traces the internal structure of the molecular cloud. 
However, it is difficult to identify extended emission associated molecular clouds due to weak emission as C$^{18}$O is less abundant compared with $^{13}$CO.  

Here we apply SCIMES to the three-dimensional data cube of the column density of $^{13}$CO ($J$=1--0) that would trace physical structures of the Orion A GMC in position-position-velocity space.  
In order to reduce computational complexity, the data is binned by 3 pixels in spatial directions before the analysis, and thus the size of a pixel is equivalent to \timeform{22.5''}.
However, this is still comparable with the effective angular resolution of the original dataset and the information of spatial structures should not be lost. 
We did not apply binning to the velocity direction to detect cloud structures that have velocity dispersions similar to the velocity resolution of 0.22 km s$^{-1}$.  
There are three parameters to input to SCIMES: {\tt min\_value}, {\tt min\_delta}, and {\tt min\_npix}.
{\tt min\_value} is the threshold of the intensity that the dendrogram uses for analyzing. A single local maximum that exceeds {\tt min\_value} is labeled as a {\it leaf}.  
{\tt min\_delta} represents the minimum height of a local maximum measured from a nearby local minimum. Only local maximums higher than {\tt min\_delta} are determined as independent {\it leaves} and these form {\it branches} with their neighbors.  If the height of a local maximum is less than {\tt min\_delta}, it just becomes a part of another {\it leaf}.
This parameter prevents identifying a local peak produced by the noise.
Both {\tt min\_value} and {\tt min\_delta} are usually given in units of $\sigma_{\mathrm{rms}}$, which corresponds to the noise level of the dataset.
{\tt min\_npix} is the minimum volume for a {\it leaf} in the position-position-velocity space that is specified by the number of pixels. 
Using the result of identification of the hierarchical structure by the dendrogram, SCIMES applies a clustering analysis to find discrete regions with similar emission properties based on graph theory. 
In this paper, we refer to the identified structure of the column density of $^{13}$CO by SCIMES as a ``cloud''. 

We adopt {\tt min\_value} = $3\sigma_{\mathrm{rms}}$, {\tt min\_delta} = $3\sigma_{\mathrm{rms}}$, and {\tt min\_npix} = 8 pixel with $\sigma_{\mathrm{rms}} = 5.61 \times 10^{13}$ cm$^{-2}$. The value of $\sigma_{\mathrm{rms}}$ corresponds to the rms noise of $0.18$ K in $T_{\mathrm{mb}}$ when we adopt the average excitation temperature of $T_{\mathrm{ex}}= 23.7$ K and the dataset achieves this noise level for most of inner parts of the mapped area \citep{Nakamura19a}. 
The distribution of the rms noise is non-uniform over the observed area. 
The value of  $\sigma_{\mathrm{rms}}$ was chosen to be suitable for detecting faint structures in the region that has relatively low rms noise, but the dendrogram can wrongly identify structures in the region that has higher rms noise.  
We evaluated the local maximum of each structure with the local rms after the identification by SCIMES and filtered it out from the catalog of clouds if the signal-to-noise ratio was less than 3.
The structures that have a peak on the map edge were also removed from the catalog as they would be artificial.
We choose {\tt min\_npix} so that identified structures contain more than 2 pixels in average along each axis, which are equivalent to \timeform{45.0''} in spatial scale and 0.44 km s$^{-1}$ in velocity.
The ``volume'' option is selected for the analysis.  

\subsection{Cloud identification}

\begin{figure*}[tbp]
\begin{center}
\FigureFile(130mm,100mm){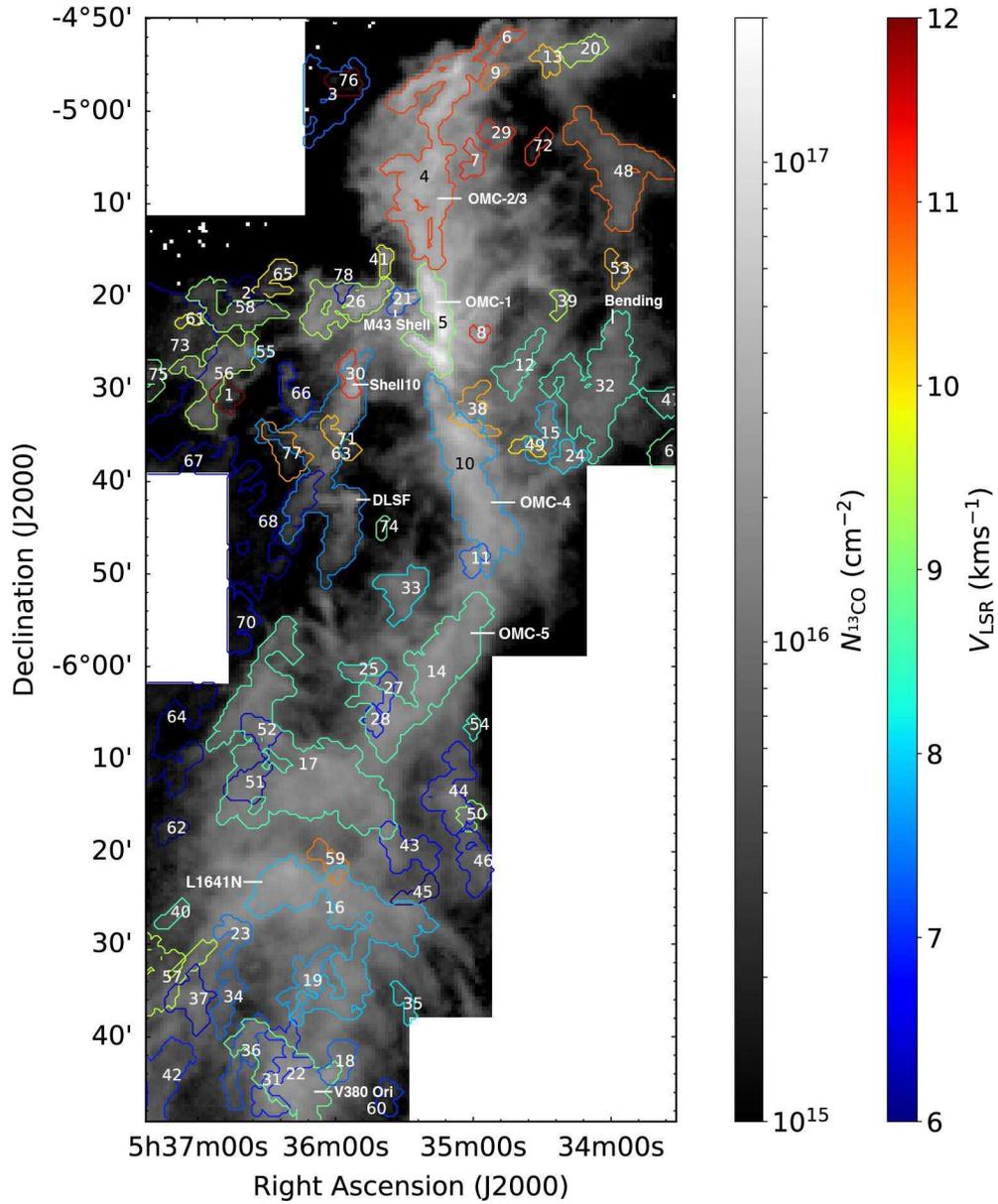}
\vspace{5mm}
\caption{Result of the identification of clouds in Orion A by SCIMES. The map of the column density of $^{13}$CO is shown in grayscale.  The numbers on clouds show the identification number of the individual clouds in this analysis. }
\label{fig:13co_scimes}
\end{center}
\end{figure*}

\begin{figure*}[tbp]
\begin{center}
\FigureFile(130mm,100mm){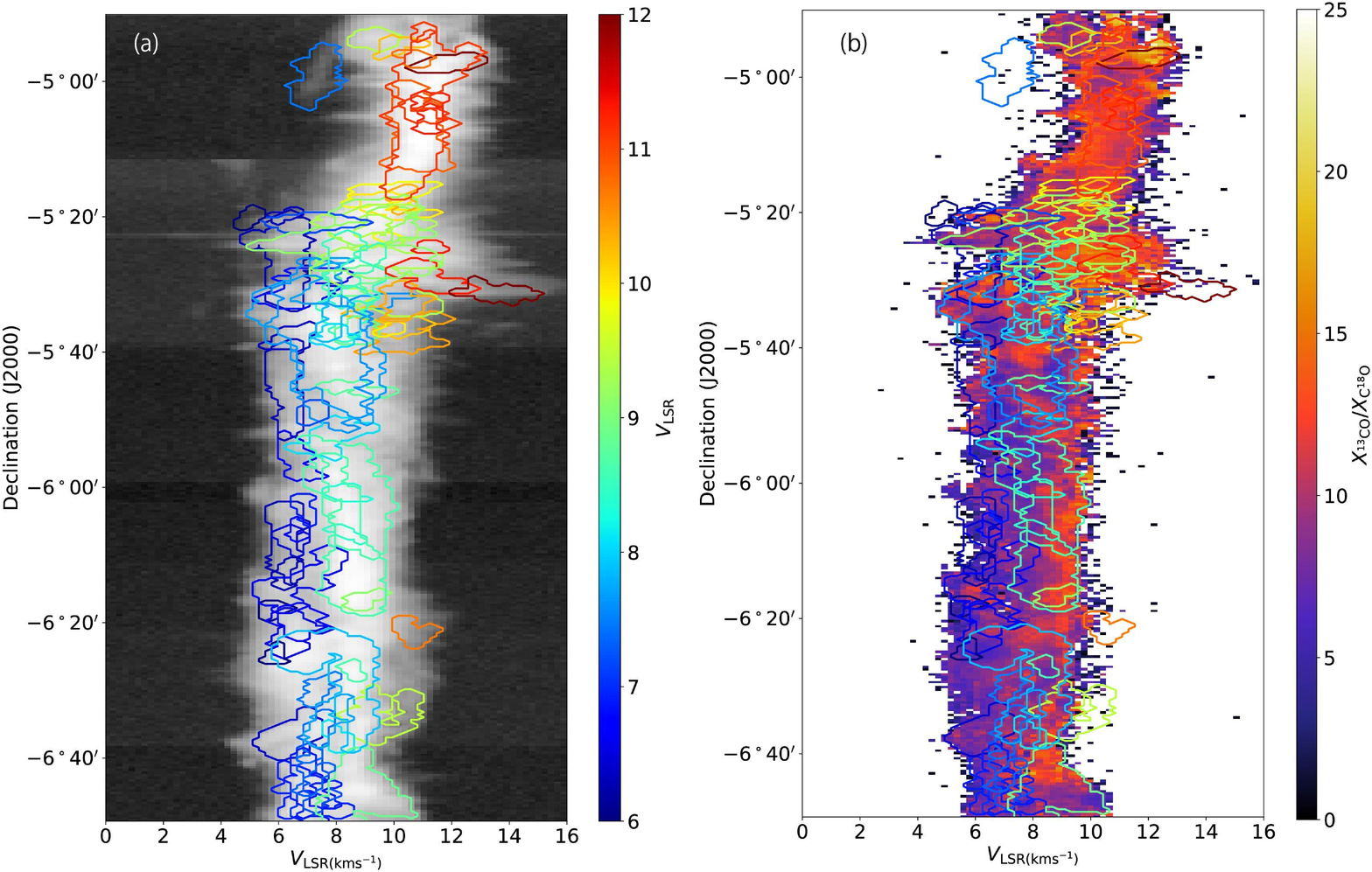}
\vspace{5mm}
\caption{Distribution of the identified clouds in the position-velocity diagram along the direction of Right Ascension. The clouds are overlaid with maps of (a) the column density of $^{13}$CO and (b) the abundance ratio of $^{13}$CO and C$^{18}$O. The clouds are outlined with a color that shows the velocity of the center of the cloud in $V_{\mathrm{LSR}}$.}
\label{fig:13co_scimes_pv}
\end{center}
\end{figure*}

Figure \ref{fig:13co_scimes} shows the spatial distribution of the final result of the identification of clouds by SCIMES with filtering.
SCIMES has identified 78 clouds including well-known structures such as OMC-1  (cloud \#5 in figure \ref{fig:13co_scimes}), OMC-2/3 (\#4), OMC-4 (\#10), OMC-5 (\#14), NGC1977 (\#6), L1641-N (\#16), and the DLSF (\#63) as summarized in table \ref{tab:scimes_clouds}.
These famous clouds are naturally identified in our analysis without any fine tuning of the input parameters. 
All clouds contain at least 2 pixels along individual axes of the data cube. 
The radius of each cloud is calculated by $R = 1.91 \sqrt{R_{\mathrm{max}} R_{\mathrm{min}} }$ as defined by \citet{Rosolowsky08}.
It is expected that a typical concentration of emission from the cloud is included within an equivalent spherical cloud that has a radius of $R$.
The rms sizes of the region, $R_{\mathrm{max}}$ and $R_{\mathrm{min}}$, are estimated from the intensity-weighted second moments along the two spatial dimensions using all pixels within the isosurface of the identified cloud. 
Most of the spatially larger clouds are located in the integral-shaped filament or the L1641-N region.  
These clouds in the integral-shaped filament are elongated along the north-south direction. In contrast, clouds in the L1641-N region tend to show elongated shapes in the east-west direction and to overlap each other. 

The other clouds are seen as emission extending to the east from Orion KL ($\alpha_{\mathrm{J2000.0}}, \delta_{\mathrm{J2000.0}} = \timeform{5h35m14.5s}, \timeform{-5D22'30.4''}$), diffuse emission at the west of the integral-shaped filament, and small structures near the edge of the map.
The clouds in diffuse emission at the west of the integral-shaped filament form a filamentary structure that connects with the northern end of the integral-shaped filament and the bending structure in the west of OMC-4.
Figure \ref{fig:13co_scimes_pv}a shows the distribution of clouds in the position-velocity diagram along the direction of right ascension with the column density map of $^{13}$CO.
A number of clouds are overlapping each other in the diagram. This indicates that clouds in the same declination have similar systemic velocity regardless of its position in right ascension.
The global velocity gradient of the emission in the integral-shaped filament \citep{Bally87} is clearly traced by clouds from the north ($V_{\mathrm{LSR}} \sim 14$ km s$^{-1}$) to the south ($\sim 8$ km s$^{-1}$) of the filament. 
The clouds in the filamentary structure in the west of the integral-shaped filament also show a gradient in velocity.
The largest velocity dispersion of a cloud is at OMC-1 that ranges from 4.2  to 11.0 km s$^{-1}$.
The most blueshifted velocity is seen at $\delta_{\mathrm{J2000.0}} = \timeform{-5D24'14.3''}$ in figure \ref{fig:13co_scimes_pv}a.
This blueshifted component of cloud \#5 traces a V-shaped profile in the position-velocity diagram that may correspond to the accelerated inflow of materials towards the Orion Nebula Cluster (ONC) from N$_2$H$^{+}$(1-0) observations \citep{Hacar17}.
The clouds extending to the east from Orion KL shows the velocity between 9 and 10 km s$^{-1}$, which is almost same with the systemic velocity of the cloud in OMC-1.
Clouds in OMC-5 and L1641-N regions have a unity velocity of $\sim 8$ km s$^{-1}$ with the dispersion of $\sim 2.5$ km s$^{-1}$.
We note that the velocities of diffuse clouds near the edge of the map are blueshifted from the main structure of Orion A. These clouds are distributed from $\delta_{\mathrm{J2000.0}} = $\timeform{-5D20'} to \timeform{-6D50'} with a system velocity of $7$ km s$^{-1}$ \citep{Sakamoto97}.

\subsection{Classification of cloud structures in Orion A}

\label{sec:classification}
\begin{figure*}[tbp]
\begin{center}
\FigureFile(130mm,100mm){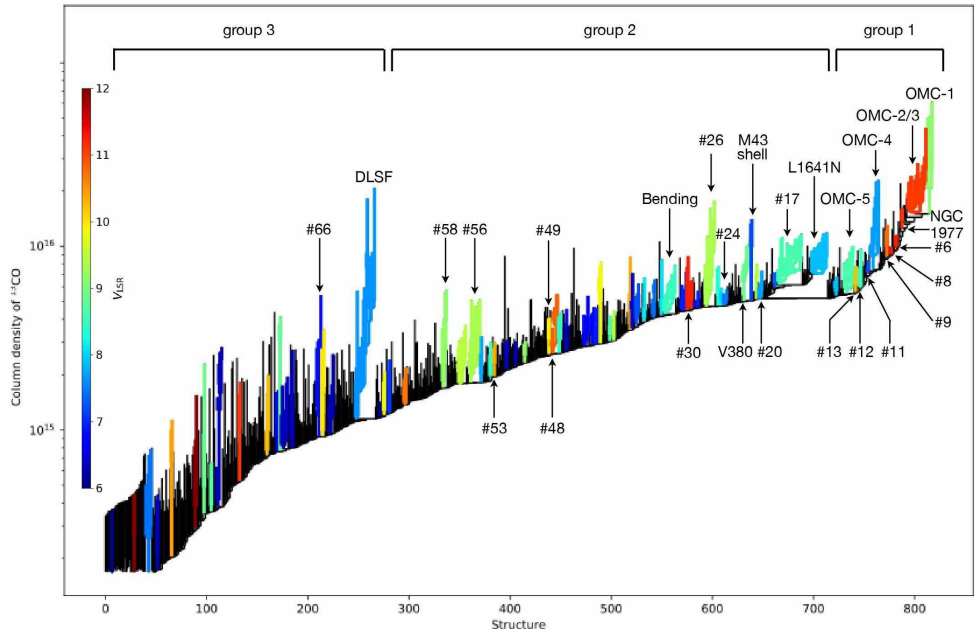}
\caption{Dendrogram of the identified clouds in Orion A. Each region outlines structures belonging to an identified cloud with a color corresponding to the velocity of a cloud. The identification numbers of structures in SCIMES are used for unnamed clouds.}
\label{fig:13co_scimes_tree}
\end{center}
\end{figure*}

The dendrogram of the analysis by SCIMES is shown in figure \ref{fig:13co_scimes_tree}. 
The color of each region in the dendrogram corresponds to the velocity of the cloud.
The leaves with $S/N < 3$ are shown in black and not regarded as clouds in this paper.
The dendrogram represents the hierarchical structures in the identified clouds in Orion A and the identified clouds can be categorized into three groups based on segregation in the dendrogram. 
The first group contains in clouds that are known as OMC-1--5 and NGC1977 in the integral-shaped filament (e.g., \cite{Odell08}). 
These clouds dominate the column density of $^{13}$CO in Orion A as we see in figure \ref{fig:COmaps}c.
The group contains several small clouds in the outer part (e.g., clouds \#8 and \#12) and the northern end (e.g., clouds \#6, \#9, and \#13) of integrated shaped filament.
A small cloud, cloud \#11, is located between OMC-4 and OMC-5 and also belongs to this group.  
Thus the integral-shaped filament and clouds in the filament are naturally separated from other parts of Orion A by SCIMES, although OMC-2 and OMC-3 are identified as a single cloud in this analysis. 
The length of a leaf indicates the range of the column density within a cloud.  
OMC-1, OMC-2/3, and OMC-4 consist of relatively long leaves compared with other clouds in this group. 
This demonstrates that these clouds have z high contrast in column density within the clouds.     

The second group includes the majority of clouds in the remaining part of Orion A, such as the southern regions of Orion A, the extended structure to the east of Ori-KL, and the filamentary structure in the west of the integral-shaped filament. 
Clouds \#16 and \#17 appear on the right-hand side of this group in the dendrogram. 
In figure \ref{fig:13co_scimes}, cloud \#16 is the molecular cloud associated with the L1641-N cluster.
Cloud \#17 is located at the southern end of the integral-shaped filament between OMC-5 and L1641-N.
Interestingly, the M43 shell shows up as an independent leaf in this group.  
Several sub-groups can be seen in the dendrogram of the second group. 
The clouds in the filamentary structure in the west of the integral-shaped filament (e.g., clouds \#20, \#24, \#32, \#48, \#49, and \#53) tend to be located in the middle of the group in figure \ref{fig:13co_scimes_tree}. 
Cloud \#32 is known as the ``bending structure'' \citep{Shimajiri11}.
In contrast, the leaves of clouds extending to the east of Ori-KL are distributed over the group.
However, some clouds in this region make a cluster on the left-hand side of the dendrogram of this group.

The third group consists of the DLSF and clouds associated with diffuse emissions in the outer part of Orion A.
The DLSF is the most prominent structure in this group and has the largest dynamic range of the column density among all the identified clouds in Orion A, which varies from $\Delta N_{^{13}\mathrm{CO}} = 1.0 \times 10^{15}$ cm$^{-2}$ to $2.0 \times 10^{16}$ cm$^{-2}$ with three peaks within the cloud. 
The second major cloud in this group, cloud \#66, is located at the northeast of the DLSF.
Most of the blueshifted clouds in figure \ref{fig:13co_scimes}b can be found in this group.

\section{Discussion}
\subsection{Origins of clouds isolated in the dendrogram}
The reason for the isolation of the DLSF from other clouds could be related to the origin of this structure.
The DLSF is thought to be formed as the result of the interaction between the expansion of the H{\scriptsize II} region in the ONC and the molecular cloud \citep{Loren79,Sugitani86,Berne14}.
The interaction can cause separation in both spatial and velocity domains. 
This is consistent with the DLSF having a large velocity gradient from the north to south \citep{Shimajiri11,Kong18}.
A similar situation can happen for the M43 shell. This is also known as the northern ionization front \citep{Berne14} and could be the result of the H{\scriptsize II} region in the ONC interacting with a foreground molecular cloud. 
In addition to DLSF and the M43 shell, an expanding shell is recently identified in the northern edge of the DLSF using $^{13}$CO data (Shell 10 in \cite{Feddersen18}). 
It is clearly seen as an isolated cloud, cloud \#30, with a C-shaped structure at 11.3 km s$^{-1}$.  
This distinct feature supports the shell having a different driving source, which is most likely to be stellar winds from T Ori as suggested by \citet{Feddersen18}. 
Thus, molecular clouds interacted with external motion would appear as isolated leaves of the dendrogram due to their prominent feature in the velocity space. 

\subsection{Relations between physical properties of the identified clouds}
Using the results of the analysis with SCIMES we derive physical properties of the identified clouds. The radius, $R$, and the full-width at half-maximum (FWHM) line width, $\Delta V$, of clouds are directly calculated by SCIMES. In addition to these parameters, the mass of clouds, $M$, is estimated by integrating the gas mass distribution in Orion A, which is converted from the map of $A_K$ following \citet{Lombardi14}.  
For clouds spatially overlapping each other, we calculated the mass associated with the individual clouds on a pixel using the weight of the column density of $^{13}$CO as follows:
\begin{equation}\label{eq:weight}
w_i (\alpha, \delta) = \frac{\Sigma_{\Delta V_i} \Delta N_{\mathrm{^{13}CO}}(\alpha, \delta, V)  }{\Sigma \Delta N_{\mathrm{^{13}CO}}(\alpha, \delta, V)},
\end{equation}
where $\Delta V_i$ is the velocity range of the $i$th cloud on a pixel and $\Delta N_{\mathrm{^{13}CO}}(\alpha, \delta, V)$ is the three-dimensional column density.

\begin{figure*}[tbp]
\begin{center}
\FigureFile(170mm,100mm){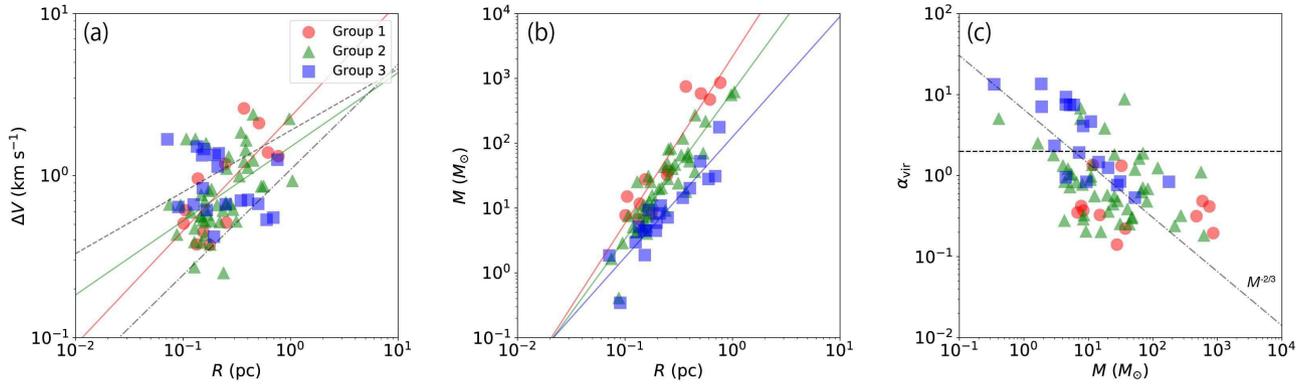}
\vspace{5mm}
\caption{Physical properties of the identified clouds by groups. (a) FMHW linewidth -- radius relation, (b) mass -- radius relation, and (c) virial parameter -- mass relation. In panels (a) and (b), red, green, and blue solid lines indicate fitting lines for clouds in group 1, 2, and, 3 respectively. Dashed and dot-dashed lines in panel (a) represent relations derived by \citet{Larson81} and \citet{Heyer04}. Dashed and dot-dashed lines in panel (c) show $\alpha_{\mathrm{vir}} = 2$ and a fitted line decreasing with $M^{-2/3}$, respectively.}
\label{fig:physical_parameter}
\end{center}
\end{figure*}

The virial ratio, the ratio of the virial mass to the mass of clouds, is obtained as 
\begin{equation}
\alpha_{\mathrm{vir}} = \frac{5 a^{-1} R \Delta V^2}{8 \ln{(2)} GM},
\end{equation}
where $a$ is a correction parameter for the mass distribution within a cloud. We assume $a = 5/3$, which corresponds to a central condensed distribution with $\rho \propto r^{-2}$ \citep{Bertoldi92}.
In table \ref{tab:clouds_properties},  we present the physical properties of 78 clouds. 
The radius of the identified clouds range from 0.07 to 1.04 pc, and the mean value is around $\sim 0.3$ pc. The cloud mass ranges from $0.3 M_{\odot}$ to $8.6 \times 10^2 M_{\odot}$. 

The relation between the line width and the radius of the cloud is widely investigated to analyze the internal motion of clouds dominated by turbulence.
Figure \ref{fig:physical_parameter}a presents the FWHM linewidth -- radius relation for the identified clouds. 
We fitted a power law to clouds in each group.
For reference, we plot relations derived by \citet{Larson81} and \citet{Heyer04} after modifications on the line width and the radius to make them consistent with the definition used here according to \citet{Maruta10}. 
The best-fitting lines for groups 1 and 2 are estimated to be $\Delta V = ({2.31\pm1.40})R^{0.70\pm0.21}$ and $(1.51\pm1.21) R^{0.46\pm0.12}$, respectively. 
The line for group 1 has a similar power index derived by \citet{Heyer04} but the coefficient is larger by a factor of 2.  
The clouds in group 2 follow Larson's relation well. 
No significant correlation is found for the clouds in group 3. 
The almost independent relation for group 3 suggests that internal motions of the clouds in this group are comparable to the inter-cloud motions as shown for $\rho$ Oph cores \citep{Maruta10}.
All the molecular clouds discussed in the previous section (DLSF, M43 shell, and cloud \#30) have line widths larger than $1$ km s$^{-1}$.
The interaction with external motion would highly enhance turbulence in these clouds.

The mass -- radius relation is plotted in figure \ref{fig:physical_parameter}b.
For a certain radius, the mass of a cloud becomes less in the order of groups 1, 2, and 3.
In other words, the density of the gas is highest for the clouds in the northern part, followed by those in the southern and outer parts of Orion A in this order.
The differences by group clearly appear in the virial parameter -- mass relation shown in figure \ref{fig:physical_parameter}c.
For $\alpha_{\mathrm{vir}} < 2$, the cloud can be self-gravitating in the absence of other forces \citep{Rosolowsky08}.
All the clouds in group 1 show virial parameters of less than 2 and it is suggested that these clouds are self-gravitating. 
The virial parameters for group 3 tend to decrease with $M^{-2/3}$, which is pointed out by \citet{Bertoldi92} for pressure-confined clumps.
For group 2, the clouds show features of both groups 1 and 3. 41 out of 47 clouds have $\alpha_{\mathrm{vir}} < 2$ and clouds tend to follow the virial parameter -- mass relation with $M^{-2/3}$ as clouds in group 3. 

\subsection{Abundance ratio of $^{13}$CO to C$^{18}$O and FUV radiation}

\begin{figure*}[tbp]
\begin{center}
\FigureFile(160mm,150mm){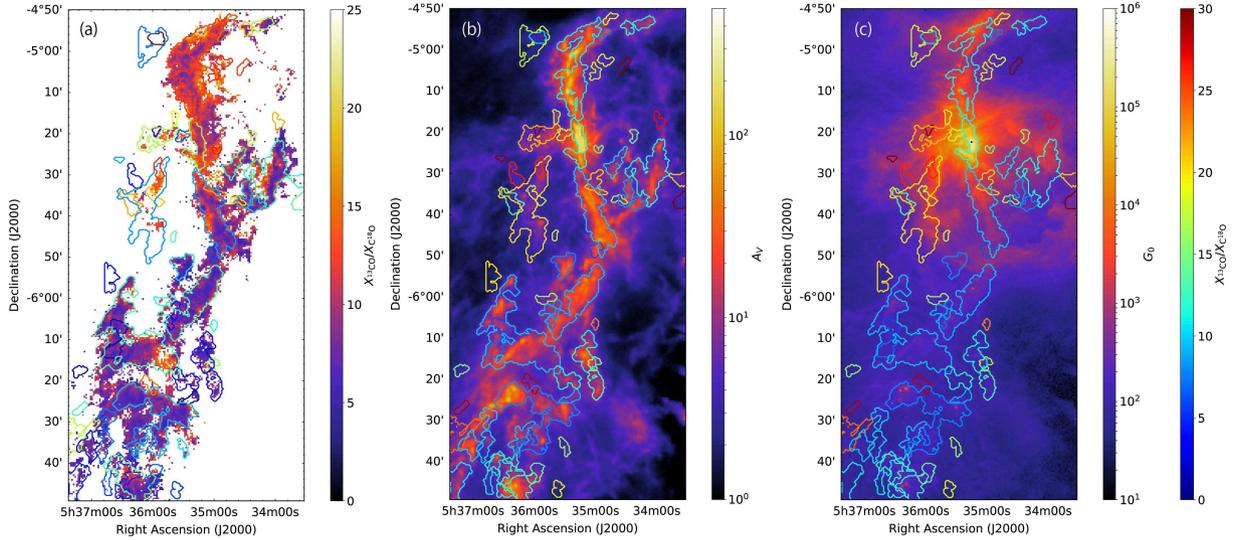}
\vspace{5mm}
\caption{Distribution of the identified clouds overlaid with maps of (a) $X_{^{13}\mathrm{CO}}/X_{\mathrm{C}^{18}\mathrm{O}}$, (b) $A_V$, and (c) $G_0$. The clouds are outlined with a color that shows the velocity of the center of the cloud in panel (a) and $X_{^{13}\mathrm{CO}}/X_{\mathrm{C}^{18}\mathrm{O}}$ in panels (b) and (c).}
\label{fig:ratiomap}
\end{center}
\end{figure*}

The abundance ratio of $^{13}$CO and C$^{18}$O, $X_{^{13}\mathrm{CO}}/X_{\mathrm{C}^{18}\mathrm{O}}$ is one of the crucial parameters that characterizes molecular clouds and their environment. 
We overlaid the identified clouds on maps of $X_{^{13}\mathrm{CO}}/X_{\mathrm{C}^{18}\mathrm{O}}$ in figure \ref{fig:ratiomap}a.
The clouds are outlined with a color that shows the median of the abundance ratio.
Each median value of the cloud reported here is computed with pixels that have a signal-to-noise ratio larger than 5.0 within the identified cloud.
As we see in subection \ref{sec:abundance}, the values of $X_{^{13}\mathrm{CO}}/X_{\mathrm{C}^{18}\mathrm{O}}$ are significantly different by clouds. 
In clouds that correspond with OMC-1, OMC-2/3, OMC-4, OMC-5, and L1641-N, the median values of the $X_{^{13}\mathrm{CO}}/X_{\mathrm{C}^{18}\mathrm{O}}$ are $11.9\pm2.1$, $11.0\pm2.8$, $9.2\pm2.3$, $7.6\pm1.6$, and $7.7\pm1.8$, respectively. 
The DLSF shows a high abundance ratio of $13.2\pm2.9$, though the data do not fully cover the entire distribution of the cloud due to weak C$^{18}$O($J=1-0$) emission.
The median of the abundance ratio of each cloud ranges from 5.6 to 17.4.
The highest ratio of $17.4\pm1.9$ is measured toward cloud \#30, which is the expanding shell (Shell 10) in \citet{Feddersen18}.
We produced the position-velocity diagram of the abundance ratio by integrating $\Delta X_{^{13}\mathrm{CO}}(\alpha, \delta, V)/\Delta X_{\mathrm{C}^{18}\mathrm{O}} (\alpha, \delta, V)$ along right ascension (see figure \ref{fig:13co_scimes_pv}b). 
In the southern part of Orion A ($\delta < \timeform{-5D40'}$), the velocity component between 8.0 and 10.0 km s$^{-1}$ shows high values of the ratio. 
In contrast, the ratio is $\sim 7$ on average for the blueshifted component of the gas. However, it is still high compared to the typical value of 5.5 in the solar system  \citep{Wilson92}.     

\begin{figure*}[tbp]
\begin{center}
\FigureFile(150mm,150mm){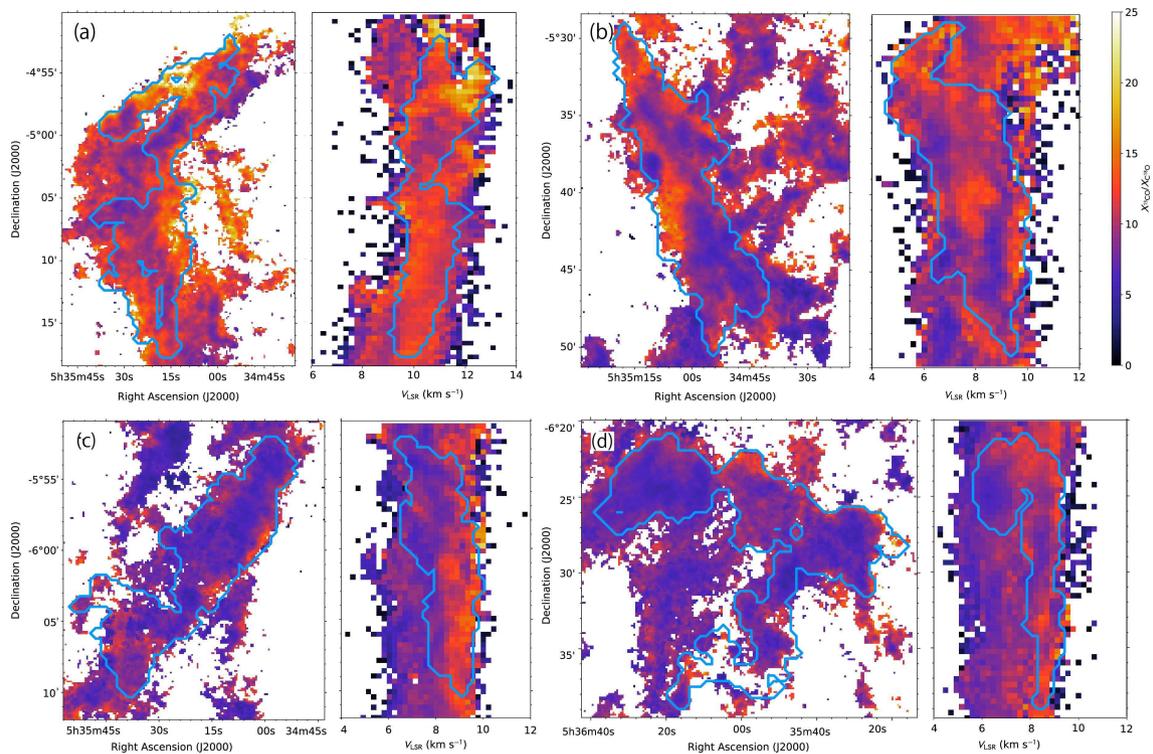}
\vspace{5mm}
\caption{Distribution of $X_{^{13}\mathrm{CO}}/X_{\mathrm{C}^{18}\mathrm{O}}$ within the identified clouds toward (a) OMC-2/3, (b) OMC-4, (c) OMC-5, and (d) L1641-N. The spatial map and the position-velocity diagram are the same as shown in figure 1f and figure 4b, but are enlarged to the clouds that are outlined by blue lines.}
\label{fig:ratiomap_OMC4_L1641N}
\end{center}
\end{figure*}

We should note that the distribution of the ratio has a large deviation even within a cloud. 
Figure \ref{fig:ratiomap_OMC4_L1641N} represents the distribution of $X_{^{13}\mathrm{CO}}/X_{\mathrm{C}^{18}\mathrm{O}}$ toward the identified clouds OMC-2/3, OMC-4, OMC-5, and L1641-N regions in the spatial maps and the position-velocity diagrams.
In these regions, the ratio is high along the edge of the cloud clearly and it decreases in the inner parts of the clouds. 
This implies that the FUV field effects at least on a scale of individual cloud structures ($\sim$ 0.3 pc) outlined with the column density of $^{13}$CO.
The northern part of OMC-2/3 shows high values in both of $X_{^{13}\mathrm{CO}}/X_{\mathrm{C}^{18}\mathrm{O}}$ and $\Delta X_{^{13}\mathrm{CO}}/\Delta  X_{\mathrm{C}^{18}\mathrm{O}}$ (see subsection \ref{sec:abundance}).
This area also presents a high CO ($J$=4--3) to CO ($J$=1--0) ratio that may be caused by interaction with OB stars or accumulation of diffuse CO components \citep{Shimajiri11, Ishii16}.
The abundance ratio also tends to be high on the edge of clouds in the position-velocity diagrams as shown in figure \ref{fig:ratiomap_OMC4_L1641N}. 
For example, the ratio exceeds 15 on the northern edge of OMC-2/3 at $(\delta_{\mathrm{J2000.0}}, V_{\mathrm{LSR}}) = (\timeform{-4D54'22''}$, $12.4$ km s$^{-1})$, the northeast edge of OMC-4 at $(\timeform{-5D32'}$, $5.2$ km s$^{-1})$, and redshift ($V_{\mathrm{LSR}} = 8 - 10$ km s$^{-1}$) components of OMC-5 and L1641-N.

\begin{figure*}[tbp]
\begin{center}
\FigureFile(150mm,150mm){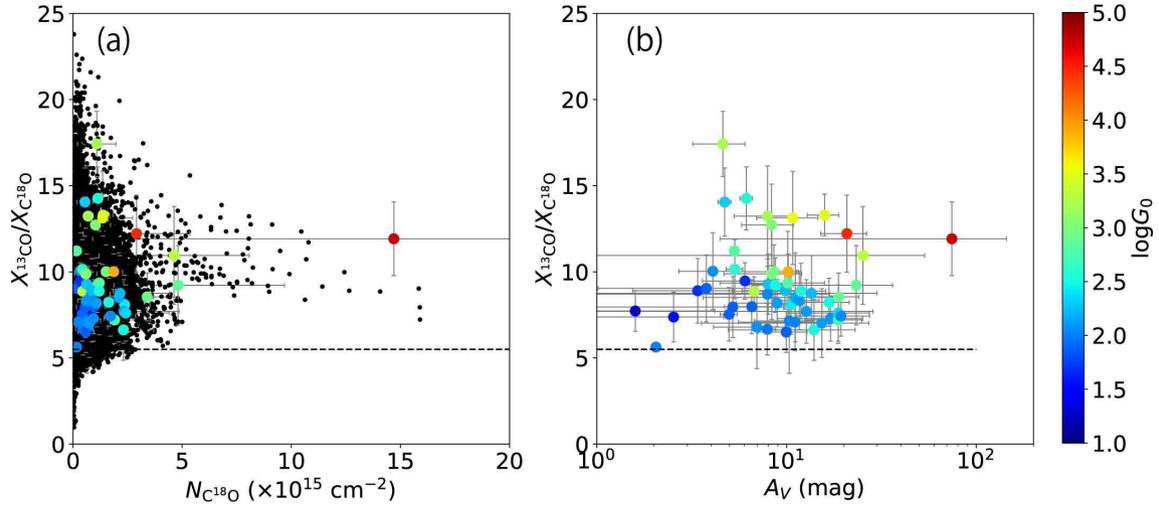}
\vspace{5mm}
\caption{Correlation plots of $X_{\mathrm{C^{13}O}}/X_{\mathrm{^{18}CO}}$ with (a) the column density of C$^{18}$O $N_{\mathrm{C^{18}O}}$ and (b) the visual extinction $A_\mathrm{V}$ for the identified clouds. Each circle represents a cloud with a color that shows the intensity of the FUV radiation field and all parameters are median in the cloud.  The error bars indicate the standard deviation of each parameter. Black dots in panel (a) represent pixel-by-pixel comparison without averaging by clouds as shown in figure \ref{fig:x13x18}b, but the data is binned by 3 pixels in spatial directions for SCIMES analysis.}
\label{fig:correlation}
\end{center}
\end{figure*}

These variations of the abundance ratio by clouds and within a cloud support the relation between the chemical difference and the PDR led by the selective photodissociation of C$^{18}$O.
Figures \ref{fig:ratiomap}b and \ref{fig:ratiomap}c show the clouds overlaid with the maps of $A_V$ and the intensity of the FUV radiation field $G_0$ estimated from infrared continuum data as described in subsection \ref{sec:dust}.   
We derived properties of $X_{^{13}\mathrm{CO}}/X_{\mathrm{C}^{18}\mathrm{O}}$,  $A_V$, and $G_0$ for the individual clouds.
Figures \ref{fig:correlation}a and \ref{fig:correlation}b represent correlations of $N_{\mathrm{C^{18}O}}$ --  $X_{^{13}\mathrm{CO}}/X_{\mathrm{C}^{18}\mathrm{O}}$ and $A_V$ -- $X_{^{13}\mathrm{CO}}/X_{\mathrm{C}^{18}\mathrm{O}}$, respectively. 
The circles and error bars indicate the median and the standard deviation for the cloud with colors that correspond to the median intensity of the FUV radiation field. 
Clouds \#71 and \#78 were flagged in the plots because the estimated values of $A_V$ and $G_0$ for these clouds have large uncertainty due to the effect of overlap with other large and dense clouds.  
In figure \ref{fig:correlation}a, we overlaid the distribution of $N_{\mathrm{C^{18}O}}$ --  $X_{^{13}\mathrm{CO}}/X_{\mathrm{C}^{18}\mathrm{O}}$ without averaging by clouds for reference. 
This is the same as the one shown in figure \ref{fig:x13x18}b, but the data is binned by 3 pixels in spatial directions for SCIMES analysis.
The values of $A_V$ and $G_0$ for clouds spatially overlapping each other are derived in the same way as the cloud mass using the weight calculated by equation (\ref{eq:weight}).
Maps of $A_V$ and $G_0$ are calculated only in the pixels where $X_{^{13}\mathrm{CO}}/X_{\mathrm{C}^{18}\mathrm{O}}$ is measured.

At low column densities ($N_{\mathrm{C^{18}O}} \lesssim 5 \times 10^{15}$ cm$^{-2}$), the median values of $X_{^{13}\mathrm{CO}}/X_{\mathrm{C}^{18}\mathrm{O}}$ for clouds show a large variation ranging from 5.6 to 17.4 as shown in figure \ref{fig:correlation}a. 
In figure \ref{fig:correlation}b, it seems that there are two trends of distribution. 
One is a set of clouds that has relatively low abundance ratios ($X_{^{13}\mathrm{CO}}/X_{\mathrm{C}^{18}\mathrm{O}} \lesssim 10$) with low column densities ($N_{\mathrm{C^{18}O}} \lesssim  5 \times 10^{15}$ cm$^{-2}$) and the other is a set of clouds that have abundance ratios higher than 10.
The low-$X_{^{13}\mathrm{CO}}/X_{\mathrm{C}^{18}\mathrm{O}}$ clouds range from 1 mag to 3 mag in $A_\mathrm{V}$ and there is no clear correlation between $A_\mathrm{V}$ and $X_{^{13}\mathrm{CO}}/X_{\mathrm{C}^{18}\mathrm{O}}$. 
The intensity of FUV for these clouds is $G_0 \lesssim 10^3$. This means that the abundance ratios are almost independent of the visual extinction of the clouds under weak FUV environment.
In contrast, the high-$X_{^{13}\mathrm{CO}}/X_{\mathrm{C}^{18}\mathrm{O}}$ clouds are located in the upper edge of the distribution in the plot and the abundance ratios decrease with the column densities at $N_{\mathrm{C^{18}O}} \gtrsim 1 \times 10^{15}$ cm$^{-2}$.  
Most of the high-$X_{^{13}\mathrm{CO}}/X_{\mathrm{C}^{18}\mathrm{O}}$ clouds are located in the integral-shaped filament and are irradiated by strong FUV of $G_0 \gtrsim 10^3$. 
In figure \ref{fig:G0correlation}, we plot $G_0$ versus $X_{^{13}\mathrm{CO}}/X_{\mathrm{C}^{18}\mathrm{O}}$ for clouds with colors that correspond to the median values of $A_\mathrm{V}$. 
This plot clearly shows that the abundance ratio increases with the intensity of the FUV radiation field and decreases with the visual extinction.  
These results are consistent with the abundance ratio being significantly affected by the selective photodissociation of C$^{18}$O, which is dominant compared to the dissociation of $^{13}$CO under a strong FUV environment and is suppressed in the dense part of the clouds. The selective photodissociation of C$^{18}$O also explains the high abundance ratio on the outer part of each cloud which is irradiated by the strong FUV and has a low column density.

\subsection{Assumptions on derivations of the abundance ratio and the FUV radiation field}
We examine the impact of assumptions used for deriving the abundance ratio and the FUV radiation field.
\subsubsection{The assumptions of the excitation temperature and the beam filling factor}
It has to be mentioned that optical depths and column densities of $^{13}$CO and C$^{18}$O were derived based on several assumptions.
$T_{\mathrm{ex}}$ is estimated by the peak temperature of $^{12}$CO. 
The excitation temperatures of $^{13}$CO ($J$=1--0) and C$^{18}$O ($J$=1--0) are likely to be lower than that of $^{12}$CO ($J$=1--0) since the result of these isotopologic lines trace inner parts of the cloud \citep{Castets90}.
As investigated by \citet{Shimajiri14}, this possibility of overestimation of $T_{\mathrm{ex}}$ has impacts for the absolute values of $X_{^{13}\mathrm{CO}}/X_{\mathrm{C}^{18}\mathrm{O}}$, but our discussion on the relative values of $X_{^{13}\mathrm{CO}}/X_{\mathrm{C}^{18}\mathrm{O}}$ should be still reliable. 

\begin{figure}[tbp]
\begin{center}
\FigureFile(80mm,80mm){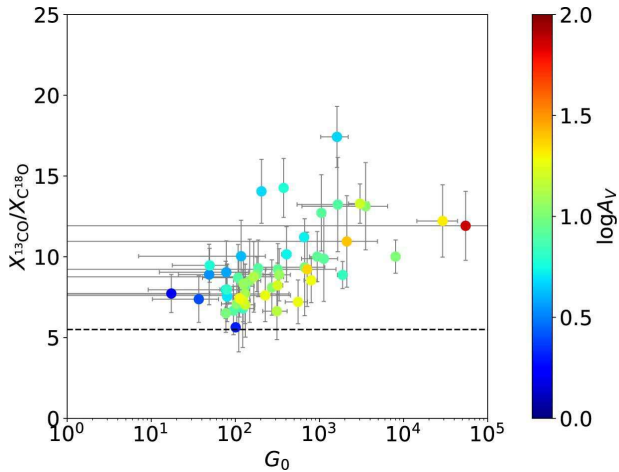}
\vspace{5mm}
\caption{Correlation plot of $X_{\mathrm{C^{13}O}}/X_{\mathrm{^{18}CO}}$ with the intensity of the FUV radiation field  $G_\mathrm{0}$ for the identified clouds. Each circle represents a cloud with a color that shows the intensity of the FUV radiation field and all parameters are median in the cloud.  The error bars indicate the standard deviation of each parameter.}
\label{fig:G0correlation}
\end{center}
\end{figure}

The beam filling factors are assumed to be 1.0 for the emission of $^{13}$CO ($J$=1--0) and C$^{18}$O ($J$=1--0).  
The distribution of C$^{18}$O ($J$=1--0) is more concentrated on dense filaments and cores than that of $^{13}$CO ($J$=1--0) as presented in figures \ref{fig:COmaps}c and \ref{fig:COmaps}e. 
This might lead to small beam filling factors of C$^{18}$O ($J$=1--0), $\phi_{\mathrm{C^{18}O}}$, of less than 1.0.
The beam size of the $^{13}$CO and C$^{18}$O data is \timeform{21.5''}, which corresponds to 0.04 pc at the distance of Orion A.
\citet{Shimajiri15} revealed that the mean size of dense cores traced by C$^{18}$O ($J$ = 1--0) in Orion A is $0.09 \pm 0.03$ pc and this gives $\phi_{\mathrm{C^{18}O}} = 0.09^2/(0.09^2+0.04^2) = 0.84$.
We investigate this effect on $X_{^{13}\mathrm{CO}}/X_{\mathrm{C}^{18}\mathrm{O}}$ using $\phi_{\mathrm{^{13}CO}} = 1.0$ and $\phi_{\mathrm{C^{18}O}} = 0.84$ as shown in figure \ref{fig:effect}a. 
$X_{^{13}\mathrm{CO}}/X_{\mathrm{C}^{18}\mathrm{O}}$ systematically decreases for the same $N_{\mathrm{C}^{18}\mathrm{O}}$ compared with the result in figure \ref{fig:x13x18}b.
The mean value of $X_{^{13}\mathrm{CO}}/X_{\mathrm{C}^{18}\mathrm{O}}$ is evaluated to be $7.76 \pm 2.21$, which is 81\% of the mean value when $\phi_{\mathrm{C^{18}O}}=1.0$ is assumed.  
However, there is no change in the relative values of  $X_{^{13}\mathrm{CO}}/X_{\mathrm{C}^{18}\mathrm{O}}$ and the mean value is 1.4 times larger than the solar system value. 
We confirmed that the variation of the abundance ratio of $X_{^{13}\mathrm{CO}}/X_{\mathrm{C}^{18}\mathrm{O}}$ is significant even after taking into account the possibility of the small beam filling factor.

\subsubsection{The omission of the 100 $\mu$m data on the estimation of the FUV radiation field}
\begin{figure*}[tbp]
\begin{center}
\FigureFile(120mm,75mm){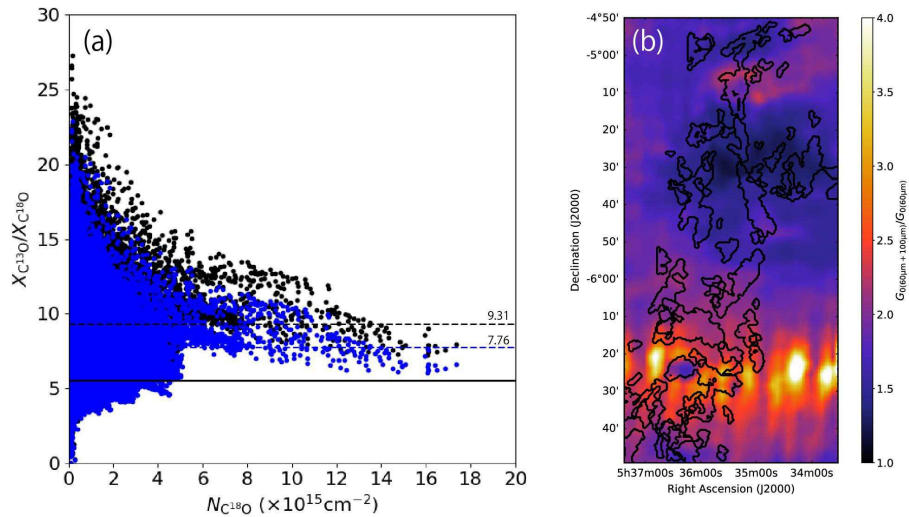}
\vspace{5mm}
\caption{(a) Effect by the small beam filling factor of the C$^{18}$O ($J$=1--0) line on the relation between $N_{\mathrm{C}^{18}\mathrm{O}}$ and $X_{^{13}\mathrm{CO}}/X_{\mathrm{C}^{18}\mathrm{O}}$. The black and blue dots present values for $\phi_{\mathrm{C^{18}O}} = 1.0$ and $0.84$, respectively. The black and blue dashed lines indicate the mean values of 9.31 and 7.76 for $\phi_{\mathrm{C^{18}O}} = 1.0$ and $0.84$. (b) Map of the ratio of the radiation field estimated from the 60 $\mu$m and 100 $\mu$m maps by IRAS, $G_{\mathrm{(60 \mu m + 100 \mu m)}}/G_{\mathrm{60 \mu m}}$. The identified clouds are presented with black lines. Note that high ratio spots at $\delta = $\timeform{-6D23'} are due to an artificial pattern in the 100 $\mu$m map.}
\label{fig:effect}
\end{center}
\end{figure*}

The lack of PACS dust continuum data at 100 $\mu$m might lead to an underestimation of the FUV radiation field. We estimate the effect of this omission using dust continuum maps at 60 $\mu$m and 100 $\mu$m taken by {\it IRAS}. 
Figure \ref{fig:effect}b represents the map of the ratio of the radiation field estimated by the 60 $\mu$m map and the 60 $\mu$m + 100 $\mu$m maps, $r = G_{\mathrm{(60 \mu m + 100 \mu m)}}/G_{\mathrm{60 \mu m}}$. The value of $r$ ranges from 1.2 to 1.6 in regions with strong FUV field. In weak FUV regions such as the southern part of Orion A and OMC-2, $r$ reaches 2.2. Since the mean values of $G_0$ for each cloud range over 3 orders of magnitude (see figure \ref{fig:G0correlation} and table \ref{tab:clouds_properties}), we conclude that the impact of the omission is small.

\section{Summary}

\label{sec:summary}
We summarize the main results of the present paper as follows.

\begin{itemize}
\item[1.] We derived wide-field distributions of the optical depths and the column densities of $^{13}$CO ($J$=1--0) and C$^{18}$O ($J$=1--0) toward the Orion A giant molecular cloud assuming LTE at the angular resolution of \timeform{22.0''} ($\approx 0.043$ pc) and velocity resolution of 0.22 km s$^{-1}$. 
The mean and maximum values of the optical depths of $^{13}$CO ($J$=1--0) and C$^{18}$O($J$=1--0) are estimated to be $\tau_{\mathrm{^{13}CO}} =0.82$ and $\tau_{\mathrm{C^{18}O}} = 5.7 \times 10^{-2}$, respectively. The mean values of column densities of the $^{13}$CO and C$^{18}$O are estimated to be $N_{\mathrm{^{13}CO}} =  6.6 \times 10^{15}$ cm$^{-2}$ and $N_{\mathrm{C^{18}O}} = 7.7 \times 10^{14}$  cm$^{-2}$, respectively.
\item[2.] The abundance ratio of $^{13}$CO and C$^{18}$O, $X_{^{13}\mathrm{CO}}/X_{\mathrm{C}^{18}\mathrm{O}}$, is estimated to be $9.31 \pm 2.65$ on average with a variation from $\sim 5$ to 27.3. The distribution of the abundance ratio in the northern part of Orion A is consistent with previous observations by \citet{Shimajiri14}.
\item[3.] We identified 78 molecular clouds from a three-dimensional data cube of the column density of $^{13}$CO using SCIMES. 
Well-known internal clouds such as OMC-1, OMC-2/3, OMC-4, OMC-5, NGC1977, L1641-N, and the DLSF are naturally identified as distinct structures in Orion A without fine-tuning the input parameters.
Identified clouds can be classified into three groups: (1) clouds in the integrated-shaped filament, (2) clouds in the southern part of Orion A and in the eastern and western sides of the integrated-shaped filament, and (3) the DLSF and clouds associated with diffuse emission.
\item[4.] We found that the DLSF, M43 shell, and Shell 10 appear as isolated structures in the dendrogram produced by SCIMES. This is because these clouds have different velocity structures from spatially-neighboring components of gas in the data cube.  This supports these clouds being likely to be produced via interactions with H{\scriptsize II} regions.  
\item[5.] The radius of the identified clouds ranges from 0.07 to 1.04 pc, and the mean value is $\sim$ 0.3 pc. The cloud mass ranges from 0.3 $M_{\odot}$ to 8.6 $\times 10^2 M_{\odot}$. The scaling relations of the physical properties show different feature between groups. The result of virial analysis shows that all clouds in group 1 are self-gravitating. The virial parameters of most of the clouds in group 3 are larger than 2 and tend to decrease with $M^{-2/3}$, which is pointed out by \citet{Bertoldi92} for pressure-confined clumps. The clouds in group 2 have features of both group 1 and group 3. 
\item[6.] The identified clouds show a significant variation in $X_{^{13}\mathrm{CO}}/X_{\mathrm{C}^{18}\mathrm{O}}$ that ranges from 5.6 to 17.4 on median over the individual clouds. Each cloud has a large deviation of the ratio within the cloud and it is also seen in the velocity direction. By combining this with the result of cloud identification, we determined that the ratio tends to be high on the edge of the cloud outlined by the column density of $^{13}$CO in the position-position-velocity space. The FUV field affects cloud structures at least on a scale of $\sim$ 0.3 pc.
\item[7.] The abundance ratio decreases from 17 to 10 with the median column density of the cloud at $N_{\mathrm{C^{18}O}} \gtrsim 1 \times 10^{15}$ cm$^{-2}$ or $A_\mathrm{V} \gtrsim 3$ mag under a strong FUV environment of $G_0 \gtrsim 10^{3}$.  The selective photodissociation of C$^{18}$O would explain these results since it is enhanced under a strong FUV environment and is suppressed in the dense part of the clouds. This is also consistent with the abundance ratio being high in the outer part of each cloud which is irradiated by the strong FUV and has a low column density.
\end{itemize}

\begin{ack}
This work was carried out as one of the large projects of the Nobeyama Radio Observatory (NRO), which is a branch of the National Astronomical 
Observatory of Japan (NAOJ), National Institute of Natural Sciences.  
We thank the NRO staff for both operating the 45 m and helping us with the data reduction.  
YS received support from the ANR (French national agency for research, project NIKA2SKY, grant agreement ANR-15-CE31-0017).
This work was supported by NAOJ ALMA Scientific Research Grant Numbers 2017-04A.
\end{ack}

\bibliographystyle{pasj}
\bibliography{orionco}

\newpage

\begin{longtable}{rllrrrrrrrl}
\caption{Clouds in Orion A identified by SCIMES}
\label{tab:scimes_clouds}
\hline 
\multicolumn{1}{c}{id}    & \multicolumn{1}{c}{$\alpha_{\mathrm{J2000}}$}  & \multicolumn{1}{c}{$\delta_{\mathrm{J2000}}$}   & \multicolumn{1}{c}{$V_\mathrm{LSR}$}        & \multicolumn{1}{c}{$R_{\mathrm{max}}$}  &\multicolumn{1}{c}{$R_{\mathrm{min}}$}   & \multicolumn{1}{c}{PA} & \multicolumn{1}{c}{$R$} & \multicolumn{1}{c}{Peak $\Delta N_{\mathrm{^{13}CO}}$} & Group & Note\\
      &                &          & \multicolumn{1}{c}{(km s$^{-1}$)}  & \multicolumn{1}{c}{(arcsec)}  & \multicolumn{1}{c}{(arcsec)}   & \multicolumn{1}{c}{(deg.)} & \multicolumn{1}{c}{(arcsec)} & \multicolumn{1}{c}{(cm$^{-2}$)} & &\\\hline
\endfirsthead
\hline 
\multicolumn{1}{c}{id}    & \multicolumn{1}{c}{$\alpha_{\mathrm{J2000}}$}  & \multicolumn{1}{c}{$\delta_{\mathrm{J2000}}$}   & \multicolumn{1}{c}{$V_\mathrm{LSR}$}        & \multicolumn{1}{c}{$R_{\mathrm{max}}$}  &\multicolumn{1}{c}{$R_{\mathrm{min}}$}   & \multicolumn{1}{c}{PA} & \multicolumn{1}{c}{$R$} & \multicolumn{1}{c}{Peak $\Delta N_{\mathrm{^{13}CO}}$} & Group & Note\\
      &                &         & \multicolumn{1}{c}{(km s$^{-1}$)}  & \multicolumn{1}{c}{(arcsec)}  & \multicolumn{1}{c}{(arcsec)}   & \multicolumn{1}{c}{(deg.)} & \multicolumn{1}{c}{(arcsec)} & \multicolumn{1}{c}{(cm$^{-2}$)} & &\\\hline
\endhead
\hline
\multicolumn{11}{l}{\footnotetext{Position angle of clouds}}
\endfoot
\hline
\endlastfoot
1	&\timeform{5h36m46.9s}	&\timeform{-5D30'51.6''}	&13.64	&869.0	&720.9	&144.5	&1511.8	&4.54$\times10^{14}$	&3	&\\
2	&\timeform{5h36m39.3s}	&\timeform{-5D19'51.1''}	&5.02	&1392.3	&925.6	&138.7	&2168.3	&3.73$\times10^{14}$	&3	&\\
3	&\timeform{5h36m1.8s}	&\timeform{-4D58'23.3''}	&7.44	&2717.0	&2046.0	&63.5	&4503.3	&7.88$\times10^{14}$	&3	&\\
4	&\timeform{5h35m18.5s}	&\timeform{-5D04'7.2''}	&11.09	&8713.7	&2345.1	&81.2	&8634.0	&4.41$\times10^{16}$	&1	&OMC-2/3\\
5	&\timeform{5h35m16.2s}	&\timeform{-5D23'3.5''}	&9.13	&3943.9	&1186.3	&95.3	&4131.4	&6.13$\times10^{16}$	&1	&OMC-1\\
6	&\timeform{5h34m46.2s}	&\timeform{-4D52'13.5''}	&11.12	&1363.8	&452.8	&214.0	&1501.0	&1.61$\times10^{16}$	&1	&NGC1977\\
7	&\timeform{5h34m59.9s}	&\timeform{-5D05'31.4''}	&11.26	&1403.0	&765.4	&85.5	&1979.3	&1.15$\times10^{16}$	&1	&\\
8	&\timeform{5h34m57.1s}	&\timeform{-5D24'9.7''}	&11.31	&685.1	&513.5	&183.2	&1132.8	&9.88$\times10^{15}$	&1	&\\
9	&\timeform{5h34m51.0s}	&\timeform{-4D56'4.5''}	&10.75	&1095.3	&345.3	&46.6	&1174.5	&1.31$\times10^{16}$	&1	&\\
10	&\timeform{5h35m2.6s}	&\timeform{-5D38'18.4''}	&7.69	&6329.9	&1407.8	&110.0	&5701.6	&2.30$\times10^{16}$	&1	&OMC-4\\
11	&\timeform{5h34m59.7s}	&\timeform{-5D48'30.3''}	&7.23	&1196.1	&686.8	&60.6	&1731.1	&8.38$\times10^{15}$	&1	&\\
12	&\timeform{5h34m40.6s}	&\timeform{-5D27'39.2''}	&8.58	&2579.1	&808.6	&56.5	&2758.3	&9.70$\times10^{15}$	&1	&\\
13	&\timeform{5h34m28.5s}	&\timeform{-4D54'21.0''}	&10.23	&1054.4	&615.4	&126.2	&1538.5	&7.82$\times10^{15}$	&1	&\\
14	&\timeform{5h35m18.9s}	&\timeform{-6D00'46.0''}	&8.64	&7389.4	&1773.0	&49.5	&6913.5	&1.00$\times10^{16}$	&1	&OMC-5\\
15	&\timeform{5h34m29.5s}	&\timeform{-5D34'59.2''}	&7.88	&2425.0	&902.8	&77.6	&2826.1	&6.86$\times10^{15}$	&1	&\\
16	&\timeform{5h36m3.3s}	&\timeform{-6D26'15.1''}	&7.89	&7294.5	&4470.7	&149.0	&10907.3	&1.19$\times10^{16}$	&2	&L1641-N\\
17	&\timeform{5h36m14.8s}	&\timeform{-6D10'43.6''}	&8.62	&7898.6	&4719.5	&134.5	&11661.6	&1.17$\times10^{16}$	&2	&\\
18	&\timeform{5h35m58.8s}	&\timeform{-6D42'50.7''}	&7.32	&1805.6	&820.2	&64.9	&2324.3	&7.24$\times10^{15}$	&2	&\\
19	&\timeform{5h36m13.0s}	&\timeform{-6D34'7.6''}	&7.73	&2531.6	&1037.3	&60.0	&3095.2	&7.36$\times10^{15}$	&2	&\\
20	&\timeform{5h34m12.1s}	&\timeform{-4D53'26.6''}	&9.40	&1604.6	&641.7	&198.3	&1938.2	&8.05$\times10^{15}$	&2	&\\
21	&\timeform{5h35m30.6s}	&\timeform{-5D20'29.0''}	&7.18	&1110.5	&524.8	&204.3	&1458.1	&1.40$\times10^{16}$	&2	&M43 shell\\
22	&\timeform{5h36m20.2s}	&\timeform{-6D44'12.9''}	&8.84	&3581.7	&1934.0	&136.4	&5026.9	&1.10$\times10^{16}$	&2	&V380 Ori\\
23	&\timeform{5h36m44.3s}	&\timeform{-6D29'4.7''}	&7.48	&1583.9	&1192.7	&74.7	&2625.2	&5.81$\times10^{15}$	&2	&\\
24	&\timeform{5h34m18.6s}	&\timeform{-5D37'27.1''}	&8.05	&1025.4	&914.8	&218.0	&1849.9	&5.94$\times10^{15}$	&2	&\\
25	&\timeform{5h35m48.4s}	&\timeform{-6D00'29.0''}	&8.36	&1722.3	&568.1	&190.8	&1889.3	&7.81$\times10^{15}$	&2	&\\
26	&\timeform{5h35m54.0s}	&\timeform{-5D20'46.4''}	&9.34	&3653.3	&1409.7	&197.5	&4334.5	&1.78$\times10^{16}$	&2	&\\
27	&\timeform{5h35m37.6s}	&\timeform{-6D02'32.6''}	&6.92	&1717.1	&445.8	&80.9	&1671.0	&5.74$\times10^{15}$	&2	&\\
28	&\timeform{5h35m43.4s}	&\timeform{-6D05'54.0''}	&6.76	&1070.8	&520.6	&98.5	&1426.1	&5.09$\times10^{15}$	&2	&\\
29	&\timeform{5h34m50.8s}	&\timeform{-5D02'32.1''}	&11.23	&1242.9	&646.3	&162.0	&1711.9	&6.52$\times10^{15}$	&2	&\\
30	&\timeform{5h35m54.2s}	&\timeform{-5D28'34.9''}	&11.26	&1809.1	&508.8	&101.1	&1832.5	&8.83$\times10^{15}$	&2	&Shell 10\\
31	&\timeform{5h36m25.5s}	&\timeform{-6D44'51.6''}	&6.92	&4122.5	&1329.4	&73.0	&4471.3	&7.28$\times10^{15}$	&2	&\\
32	&\timeform{5h34m5.6s}	&\timeform{-5D29'53.3''}	&8.49	&4432.1	&2445.5	&73.9	&6288.2	&7.91$\times10^{15}$	&2	&Bending structure\\
33	&\timeform{5h35m30.0s}	&\timeform{-5D51'44.5''}	&8.07	&1753.6	&1390.7	&55.7	&2982.7	&8.50$\times10^{15}$	&2	&\\
34	&\timeform{5h36m47.3s}	&\timeform{-6D35'51.8''}	&7.39	&2345.6	&706.2	&76.2	&2458.2	&6.61$\times10^{15}$	&2	&\\
35	&\timeform{5h35m29.2s}	&\timeform{-6D36'37.9''}	&8.11	&1550.3	&469.5	&111.7	&1629.5	&6.74$\times10^{15}$	&2	&\\
36	&\timeform{5h36m39.8s}	&\timeform{-6D41'41.9''}	&7.24	&3851.1	&873.0	&71.1	&3502.1	&6.27$\times10^{15}$	&2	&\\
37	&\timeform{5h37m2.5s}	&\timeform{-6D36'6.0''}	&6.40	&1761.0	&1255.9	&108.0	&2840.5	&7.15$\times10^{15}$	&2	&\\
38	&\timeform{5h35m1.2s}	&\timeform{-5D32'23.4''}	&10.37	&1695.6	&1271.5	&97.0	&2804.4	&8.73$\times10^{15}$	&2	&\\
39	&\timeform{5h34m22.0s}	&\timeform{-5D20'44.4''}	&9.38	&1068.2	&453.9	&84.3	&1329.9	&4.49$\times10^{15}$	&2	&\\
40	&\timeform{5h37m10.4s}	&\timeform{-6D26'43.5''}	&8.62	&1457.1	&383.6	&220.9	&1427.9	&4.18$\times10^{15}$	&2	&\\
41	&\timeform{5h35m38.2s}	&\timeform{-5D16'5.1''}	&9.85	&1071.5	&369.6	&86.4	&1201.9	&8.22$\times10^{15}$	&2	&\\
42	&\timeform{5h37m14.2s}	&\timeform{-6D44'20.5''}	&6.93	&3105.6	&1233.6	&69.6	&3738.5	&4.62$\times10^{15}$	&2	&\\
43	&\timeform{5h35m30.7s}	&\timeform{-6D19'32.7''}	&6.72	&2320.3	&1260.8	&152.9	&3266.9	&4.67$\times10^{15}$	&2	&\\
44	&\timeform{5h35m9.3s}	&\timeform{-6D13'38.8''}	&6.64	&2910.7	&1413.1	&83.7	&3873.6	&5.36$\times10^{15}$	&2	&\\
45	&\timeform{5h35m24.9s}	&\timeform{-6D24'34.6''}	&5.86	&1969.1	&559.3	&204.1	&2004.5	&3.70$\times10^{15}$	&2	&\\
46	&\timeform{5h34m58.5s}	&\timeform{-6D21'12.7''}	&6.55	&2179.5	&989.0	&102.4	&2804.1	&4.57$\times10^{15}$	&2	&\\
47	&\timeform{5h33m36.8s}	&\timeform{-5D31'23.7''}	&8.47	&1162.7	&750.1	&164.1	&1783.8	&4.42$\times10^{15}$	&2	&\\
48	&\timeform{5h33m57.6s}	&\timeform{-5D06'41.6''}	&10.85	&4240.8	&2315.2	&113.4	&5984.8	&5.47$\times10^{15}$	&2	&\\
49	&\timeform{5h34m36.1s}	&\timeform{-5D36'18.4''}	&9.95	&1222.5	&465.3	&163.6	&1440.5	&4.10$\times10^{15}$	&2	&\\
50	&\timeform{5h35m1.5s}	&\timeform{-6D16'9.8''}	&9.12	&1034.3	&722.6	&61.2	&1651.2	&3.05$\times10^{15}$	&2	&\\
51	&\timeform{5h36m37.8s}	&\timeform{-6D12'41.7''}	&6.27	&1907.4	&1021.0	&221.0	&2665.4	&3.03$\times10^{15}$	&2	&\\
52	&\timeform{5h36m32.7s}	&\timeform{-6D07'0.4''}	&6.45	&1422.9	&802.0	&141.1	&2040.4	&3.08$\times10^{15}$	&2	&\\
53	&\timeform{5h33m57.4s}	&\timeform{-5D17'1.6''}	&10.32	&1332.8	&660.6	&122.7	&1792.1	&3.00$\times10^{15}$	&2	&\\
54	&\timeform{5h35m0.1s}	&\timeform{-6D06'27.3''}	&8.44	&737.2	&419.8	&82.0	&1062.6	&3.01$\times10^{15}$	&2	&\\
55	&\timeform{5h36m33.1s}	&\timeform{-5D26'11.6''}	&7.74	&799.0	&240.2	&173.9	&836.8	&3.24$\times10^{15}$	&2	&\\
56	&\timeform{5h36m51.3s}	&\timeform{-5D28'32.4''}	&9.35	&3897.1	&1774.6	&58.1	&5022.9	&5.15$\times10^{15}$	&2	&\\
57	&\timeform{5h37m14.1s}	&\timeform{-6D33'43.4''}	&9.46	&2861.4	&1756.8	&57.3	&4282.4	&3.72$\times10^{15}$	&2	&\\
58	&\timeform{5h36m42.1s}	&\timeform{-5D21'22.9''}	&9.16	&3501.8	&1339.5	&165.8	&4136.7	&5.79$\times10^{15}$	&2	&\\
59	&\timeform{5h36m3.0s}	&\timeform{-6D20'57.6''}	&10.70	&1724.6	&675.2	&130.8	&2061.0	&2.19$\times10^{15}$	&2	&\\
60	&\timeform{5h35m37.0s}	&\timeform{-6D46'31.5''}	&7.08	&1347.7	&651.3	&84.4	&1789.5	&2.42$\times10^{15}$	&2	&\\
61	&\timeform{5h37m3.7s}	&\timeform{-5D22'37.8''}	&9.87	&920.4	&287.7	&193.2	&982.9	&2.09$\times10^{15}$	&2	&\\
62	&\timeform{5h37m12.0s}	&\timeform{-6D17'38.0''}	&5.84	&1099.0	&676.2	&207.3	&1646.6	&1.43$\times10^{15}$	&2	&\\
63	&\timeform{5h36m0.4s}	&\timeform{-5D37'13.1''}	&7.63	&7380.4	&2651.7	&85.3	&8449.6	&2.08$\times10^{16}$	&3	&DLSF\\
64	&\timeform{5h37m12.0s}	&\timeform{-6D05'38.4''}	&6.27	&4491.0	&2742.6	&82.9	&6703.2	&2.59$\times10^{15}$	&3	&\\
65	&\timeform{5h36m25.6s}	&\timeform{-5D17'49.8''}	&9.98	&1683.4	&790.7	&48.5	&2203.6	&3.52$\times10^{15}$	&3	&\\
66	&\timeform{5h36m17.8s}	&\timeform{-5D30'40.7''}	&6.84	&2155.5	&738.3	&112.6	&2409.6	&5.43$\times10^{15}$	&3	&\\
67	&\timeform{5h37m4.5s}	&\timeform{-5D38'0.1''}	&6.37	&4218.0	&973.0	&184.9	&3869.4	&2.73$\times10^{15}$	&3	&\\
68	&\timeform{5h36m32.2s}	&\timeform{-5D44'32.6''}	&6.30	&4712.8	&3487.0	&66.9	&7742.8	&1.95$\times10^{15}$	&3	&\\
69	&\timeform{5h33m35.6s}	&\timeform{-5D36'55.7''}	&8.83	&1190.3	&797.0	&78.4	&1860.3	&4.17$\times10^{15}$	&3	&\\
70	&\timeform{5h36m41.8s}	&\timeform{-5D55'26.6''}	&6.69	&2169.0	&986.2	&88.9	&2793.4	&1.65$\times10^{15}$	&3	&\\
71	&\timeform{5h35m57.9s}	&\timeform{-5D35'37.8''}	&10.22	&1751.1	&446.6	&135.1	&1689.1	&1.99$\times10^{15}$	&3	&\\
72	&\timeform{5h34m32.7s}	&\timeform{-5D03'56.7''}	&11.09	&1438.0	&377.2	&52.8	&1406.8	&1.83$\times10^{15}$	&3	&\\
73	&\timeform{5h37m10.5s}	&\timeform{-5D25'30.5''}	&6.29	&3930.9	&2188.2	&78.0	&5601.7	&2.83$\times10^{15}$	&3	&\\
74	&\timeform{5h35m39.6s}	&\timeform{-5D45'5.0''}	&8.76	&726.6	&242.5	&85.8	&801.7	&6.71$\times10^{14}$	&3	&\\
75	&\timeform{5h37m19.8s}	&\timeform{-5D28'37.7''}	&8.87	&1396.0	&574.8	&91.6	&1711.0	&2.31$\times10^{15}$	&3	&\\
76	&\timeform{5h35m57.2s}	&\timeform{-4D56'52.9''}	&12.09	&1183.5	&712.9	&173.5	&1754.4	&1.55$\times10^{15}$	&3	&\\
77	&\timeform{5h36m21.8s}	&\timeform{-5D37'9.3''}	&10.43	&1948.6	&776.1	&118.2	&2348.8	&1.14$\times10^{15}$	&3	&\\
78	&\timeform{5h35m56.5s}	&\timeform{-5D19'20.4''}	&6.35	&670.6	&418.7	&58.0	&1012.1	&4.46$\times10^{14}$	&3	&\\
\end{longtable}

\begin{longtable}{rrrrrrrrrr}
\caption{Physical Properties of Clouds}
\label{tab:clouds_properties}
\hline 
\multicolumn{1}{c}{id}    & \multicolumn{1}{c}{$R$}  & \multicolumn{1}{c}{$M$}   & \multicolumn{1}{c}{$\Delta V$}        & \multicolumn{1}{c}{$\alpha_{\mathrm{vir}}$}  &\multicolumn{1}{c}{$N_{\mathrm{^{13}CO}}$}   & \multicolumn{1}{c}{$N_{\mathrm{C^{18}O}}$} & \multicolumn{1}{c}{$X_{^{13}\mathrm{CO}}/X_{\mathrm{C}^{18}\mathrm{O}}$} & \multicolumn{1}{c}{$A_V$} & \multicolumn{1}{c}{$G_0$}\\
      &   \multicolumn{1}{c}{(pc)}     &       \multicolumn{1}{c}{($M_{\odot}$)}      & \multicolumn{1}{c}{(km s$^{-1}$)}  &   & \multicolumn{1}{c}{($\times 10^{16}$ cm$^{-2}$)}   & \multicolumn{1}{c}{($\times 10^{15}$ cm$^{-2}$)} &  & \multicolumn{1}{c}{(mag)} & \multicolumn{1}{c}{(Habing unit)} \\\hline
\endfirsthead
\hline 
\multicolumn{1}{c}{id}    & \multicolumn{1}{c}{$R$}  & \multicolumn{1}{c}{$M$}   & \multicolumn{1}{c}{$\Delta V$}        & \multicolumn{1}{c}{$\alpha_{\mathrm{vir}}$}  &\multicolumn{1}{c}{$N_{\mathrm{^{13}CO}}$}   & \multicolumn{1}{c}{$N_{\mathrm{C^{18}O}}$} & \multicolumn{1}{c}{$X_{^{13}\mathrm{CO}}/X_{\mathrm{C}^{18}\mathrm{O}}$} & \multicolumn{1}{c}{$A_V$} & \multicolumn{1}{c}{$G_0$}\\
      &   \multicolumn{1}{c}{(pc)}     &       \multicolumn{1}{c}{($M_{\odot}$)}      & \multicolumn{1}{c}{(km s$^{-1}$)}  &   & \multicolumn{1}{c}{($\times 10^{16}$ cm$^{-2}$)}   & \multicolumn{1}{c}{($\times 10^{15}$ cm$^{-2}$)} &  & \multicolumn{1}{c}{(mag)} & \multicolumn{1}{c}{(Habing unit)} \\\hline
\endhead
\hline
\multicolumn{10}{l}{\footnotetext{Position angle of clouds}}
\endfoot
\hline
\endlastfoot
1	&0.13	&5.2	&1.51	&7.46	&0.06	&-	&-	&-	&-\\
2	&0.19	&4.5	&0.42	&0.96	&0.03	&-	&-	&-	&-\\
3	&0.40	&20.2	&0.71	&1.25	&0.08	&0.06	&-	&-	&-\\
4	&0.77	&860.4	&1.32	&0.20	&5.11	&4.63	&11.0$\pm$2.8	&25.3$\pm$28.1	&2112.4$\pm$2766.8\\
5	&0.37	&750.7	&2.60	&0.42	&17.82	&14.71	&11.9$\pm$2.1	&74.3$\pm$69.9	&54546.7$\pm$86058.4\\
6	&0.13	&6.8	&0.38	&0.35	&1.59	&1.03	&12.7$\pm$2.4	&8.3$\pm$2.5	&1053.5$\pm$116.7\\
7	&0.18	&8.4	&0.37	&0.37	&2.05	&1.14	&14.3$\pm$1.8	&6.1$\pm$0.5	&374.8$\pm$23.2\\
8	&0.10	&7.7	&0.51	&0.42	&1.92	&1.38	&13.3$\pm$1.2	&15.9$\pm$3.0	&3028.6$\pm$808.6\\
9	&0.10	&15.1	&0.61	&0.33	&3.03	&3.40	&8.5$\pm$1.4	&18.8$\pm$5.3	&797.4$\pm$142.2\\
10	&0.51	&585.9	&2.11	&0.49	&4.44	&4.82	&9.2$\pm$2.3	&23.3$\pm$12.9	&713.8$\pm$858.7\\
11	&0.15	&27.7	&0.45	&0.14	&1.47	&1.91	&7.2$\pm$1.4	&18.7$\pm$6.2	&555.8$\pm$164.2\\
12	&0.25	&32.6	&1.18	&1.32	&1.37	&1.52	&10.0$\pm$1.6	&8.6$\pm$3.0	&940.0$\pm$174.7\\
13	&0.14	&11.7	&0.96	&1.36	&1.19	&1.10	&9.2$\pm$1.6	&8.7$\pm$2.1	&319.5$\pm$89.3\\
14	&0.62	&472.8	&1.39	&0.32	&2.55	&2.42	&7.6$\pm$1.6	&18.7$\pm$8.6	&226.6$\pm$226.4\\
15	&0.25	&37.4	&0.51	&0.22	&1.12	&1.20	&8.9$\pm$1.6	&11.9$\pm$4.2	&331.9$\pm$119.0\\
16	&0.97	&556.9	&2.24	&1.10	&1.72	&2.33	&7.7$\pm$1.8	&12.7$\pm$15.9	&130.4$\pm$304.0\\
17	&1.04	&617.0	&0.93	&0.18	&2.23	&2.10	&8.5$\pm$2.6	&11.1$\pm$7.5	&147.9$\pm$52.6\\
18	&0.21	&19.6	&0.66	&0.57	&1.20	&0.69	&6.8$\pm$2.4	&7.0$\pm$2.9	&122.5$\pm$47.1\\
19	&0.28	&40.3	&0.67	&0.39	&1.11	&1.08	&8.2$\pm$1.4	&8.9$\pm$3.5	&130.9$\pm$36.7\\
20	&0.17	&25.3	&0.81	&0.56	&2.76	&2.37	&8.1$\pm$1.7	&10.6$\pm$1.6	&271.8$\pm$27.4\\
21	&0.13	&25.3	&1.68	&1.82	&5.60	&2.91	&12.2$\pm$2.2	&20.8$\pm$5.6	&29070.1$\pm$14684.1\\
22	&0.45	&271.5	&1.24	&0.32	&2.44	&2.15	&8.8$\pm$2.2	&13.5$\pm$16.4	&165.6$\pm$208.9\\
23	&0.23	&43.1	&0.61	&0.25	&1.00	&0.94	&7.3$\pm$1.3	&17.1$\pm$8.1	&123.6$\pm$26.4\\
24	&0.17	&31.0	&0.62	&0.25	&1.08	&1.64	&8.2$\pm$1.4	&16.7$\pm$4.7	&317.5$\pm$48.1\\
25	&0.17	&12.5	&0.57	&0.56	&1.10	&0.63	&9.3$\pm$1.8	&8.0$\pm$1.6	&186.8$\pm$12.3\\
26	&0.39	&70.3	&1.65	&1.88	&2.89	&1.36	&13.1$\pm$2.7	&10.8$\pm$3.8	&3536.8$\pm$2918.1\\
27	&0.15	&8.5	&0.39	&0.33	&0.50	&0.86	&8.9$\pm$1.1	&9.8$\pm$4.0	&180.2$\pm$79.5\\
28	&0.13	&4.2	&0.27	&0.28	&0.49	&0.91	&8.9$\pm$1.9	&3.4$\pm$1.9	&48.8$\pm$27.8\\
29	&0.15	&6.5	&0.53	&0.83	&1.07	&0.55	&14.1$\pm$2.0	&4.7$\pm$0.4	&204.4$\pm$21.0\\
30	&0.16	&7.5	&1.58	&6.80	&2.86	&1.09	&17.4$\pm$1.9	&4.6$\pm$1.4	&1615.4$\pm$581.0\\
31	&0.40	&52.8	&1.12	&1.19	&0.97	&0.85	&8.7$\pm$2.1	&7.9$\pm$9.0	&107.9$\pm$67.1\\
32	&0.56	&218.0	&0.86	&0.24	&1.48	&1.21	&9.4$\pm$3.0	&10.1$\pm$5.7	&666.8$\pm$490.0\\
33	&0.27	&83.5	&1.31	&0.68	&1.93	&2.31	&6.6$\pm$1.8	&14.0$\pm$9.0	&311.9$\pm$100.2\\
34	&0.22	&20.4	&0.52	&0.36	&0.98	&0.95	&8.0$\pm$1.6	&6.6$\pm$1.8	&79.6$\pm$14.1\\
35	&0.15	&5.5	&0.56	&1.04	&0.85	&0.43	&7.5$\pm$1.6	&5.0$\pm$0.5	&79.9$\pm$7.2\\
36	&0.31	&24.1	&0.52	&0.44	&0.63	&0.57	&8.0$\pm$1.8	&5.2$\pm$4.5	&76.9$\pm$67.7\\
37	&0.25	&80.5	&1.10	&0.48	&1.74	&1.71	&7.4$\pm$1.2	&19.3$\pm$8.0	&112.0$\pm$18.7\\
38	&0.25	&28.3	&0.66	&0.49	&1.30	&0.59	&9.9$\pm$2.6	&8.3$\pm$3.7	&1126.7$\pm$514.6\\
39	&0.12	&5.1	&0.59	&1.03	&0.67	&0.45	&8.9$\pm$0.8	&6.7$\pm$1.6	&1881.8$\pm$250.6\\
40	&0.13	&8.3	&0.39	&0.29	&0.51	&0.16	&-	&-	&-\\
41	&0.11	&7.5	&1.68	&5.04	&3.16	&1.86	&10.0$\pm$1.0	&10.2$\pm$1.1	&8032.8$\pm$913.9\\
42	&0.33	&65.6	&0.99	&0.63	&0.67	&0.59	&6.5$\pm$1.2	&10.0$\pm$2.9	&77.2$\pm$9.6\\
43	&0.29	&46.7	&0.62	&0.30	&0.72	&0.66	&7.1$\pm$3.0	&10.3$\pm$3.6	&109.6$\pm$26.5\\
44	&0.35	&119.0	&1.85	&1.24	&0.93	&0.75	&7.0$\pm$2.0	&15.3$\pm$11.8	&131.6$\pm$39.1\\
45	&0.18	&15.6	&0.38	&0.20	&0.32	&0.46	&6.7$\pm$1.5	&7.9$\pm$0.8	&95.3$\pm$27.4\\
46	&0.25	&47.7	&0.68	&0.31	&0.92	&0.62	&8.3$\pm$1.7	&11.7$\pm$3.2	&124.8$\pm$15.5\\
47	&0.16	&10.8	&0.69	&0.89	&0.72	&0.43	&10.2$\pm$1.7	&5.3$\pm$0.5	&406.3$\pm$51.9\\
48	&0.53	&69.8	&0.86	&0.71	&1.15	&-	&-	&-	&-\\
49	&0.13	&4.3	&0.47	&0.84	&0.58	&0.36	&10.0$\pm$2.2	&4.1$\pm$1.4	&117.1$\pm$110.0\\
50	&0.15	&11.0	&0.76	&0.97	&0.51	&0.33	&9.5$\pm$1.1	&6.0$\pm$2.4	&49.5$\pm$31.7\\
51	&0.24	&9.2	&0.25	&0.20	&0.26	&0.41	&7.7$\pm$1.2	&1.6$\pm$2.6	&17.3$\pm$32.2\\
52	&0.18	&8.1	&0.52	&0.78	&0.49	&0.86	&7.4$\pm$1.4	&2.5$\pm$2.1	&36.8$\pm$26.4\\
53	&0.16	&5.2	&0.43	&0.72	&0.45	&0.16	&11.2$\pm$0.0	&5.3$\pm$0.0	&656.8$\pm$0.0\\
54	&0.09	&2.9	&0.66	&1.79	&0.44	&0.22	&-	&-	&-\\
55	&0.07	&1.6	&0.66	&2.51	&0.44	&0.18	&-	&-	&-\\
56	&0.45	&36.3	&2.39	&8.81	&1.54	&-	&-	&-	&-\\
57	&0.38	&61.2	&1.44	&1.62	&0.47	&0.24	&7.1$\pm$1.6	&11.1$\pm$5.5	&103.9$\pm$35.5\\
58	&0.37	&17.9	&1.22	&3.84	&0.76	&-	&-	&-	&-\\
59	&0.18	&11.5	&0.82	&1.36	&0.33	&0.16	&9.0$\pm$1.9	&3.8$\pm$2.8	&78.1$\pm$65.6\\
60	&0.16	&4.0	&0.52	&1.33	&0.31	&0.20	&-	&-	&-\\
61	&0.09	&0.4	&0.43	&5.05	&0.28	&-	&-	&-	&-\\
62	&0.15	&4.8	&0.56	&1.21	&0.24	&0.13	&-	&-	&-\\
63	&0.75	&176.9	&1.25	&0.84	&1.50	&0.69	&13.2$\pm$2.9	&7.9$\pm$2.6	&1651.0$\pm$1109.7\\
64	&0.60	&28.1	&0.53	&0.76	&0.23	&-	&-	&-	&-\\
65	&0.20	&5.9	&1.34	&7.48	&0.50	&-	&-	&-	&-\\
66	&0.21	&10.9	&1.37	&4.65	&0.88	&0.31	&-	&-	&-\\
67	&0.35	&14.4	&0.70	&1.47	&0.25	&-	&-	&-	&-\\
68	&0.69	&31.0	&0.55	&0.85	&0.18	&-	&-	&-	&-\\
69	&0.17	&9.3	&0.61	&0.84	&0.57	&0.19	&-	&-	&-\\
70	&0.25	&7.2	&0.67	&1.93	&0.25	&0.16	&5.6$\pm$0.0	&2.0$\pm$0.0	&100.8$\pm$0.0\\
71	&0.15	&4.5	&1.34	&7.54	&0.34	&0.20	&14.4$\pm$2.0	&0.2$\pm$0.7	&36.7$\pm$120.0\\
72	&0.13	&3.0	&0.66	&2.33	&0.27	&0.09	&-	&-	&-\\
73	&0.50	&52.5	&0.67	&0.54	&0.30	&-	&-	&-	&-\\
74	&0.07	&1.9	&1.68	&13.58	&0.14	&0.13	&-	&-	&-\\
75	&0.15	&1.9	&0.84	&7.10	&0.21	&-	&-	&-	&-\\
76	&0.16	&4.5	&1.46	&9.34	&0.18	&0.20	&-	&-	&-\\
77	&0.21	&8.3	&1.14	&4.10	&0.14	&0.15	&-	&-	&-\\
78	&0.09	&0.3	&0.64	&13.44	&0.04	&0.12	&9.3$\pm$0.0	&0.1$\pm$0.0	&15.5$\pm$0.0\\
\end{longtable}

\end{document}